  \documentclass[12pt,final,onecolumn]{IEEEtran}

\usepackage{graphicx}
\usepackage{color}

\usepackage{amsthm,thmtools}

\usepackage{setspace}
\usepackage{enumerate}
\usepackage{verbatim}
\usepackage{amssymb}
\usepackage{amsbsy}

\usepackage{url}

\usepackage[dvips]{epsfig}

\usepackage{amsmath}
\newcommand*\diff{\mathop{}\!\mathrm{d}}

\def\urltilde{\kern -.15em\lower .7ex\hbox{\~{}}\kern .04em}
\def\urldot{\kern -.10em.\kern -.10em}

\def\urlhttp{http\kern -.10em\lower -.1ex\hbox{:}\kern -.12em\lower 0ex\hbox{/}\kern -.18em\lower 0ex\hbox{/}}

\theoremstyle{plain}

\declaretheorem[name={Theorem}  ] {Theorem}

\declaretheorem[name={Problem}, sibling=Theorem] {Problem}
\declaretheorem[name={Assumption}, sibling=Theorem] {Assumption}

\declaretheorem[name={Lemma}, sibling=Theorem] {Lemma}
\declaretheorem[name={Proposition}, sibling=Theorem] {Proposition}
\declaretheorem[name={Corollary}, sibling=Theorem] {Corollary}
\declaretheorem[name={Example},qed={\lower-0.3ex\hbox{$\square$}}, sibling=Theorem] {Example}

\newcommand {\R}{\mathbb R}

\newcommand{\be}{\begin{equation}}
\newcommand{\ee}{\end{equation}}
\newcommand{\inter}{\operatorname{{\mathrm relint}}}
\newcommand{\interior}{\operatorname{{\mathrm int}}}

\newcommand\mycom[2]{\genfrac{}{}{0pt}{}{#1}{#2}}

\newcommand {\N}{\mathbb N}

\newcommand{\Ac}{\mathcal A}
\newcommand{\BB}{\mathcal B}
\newcommand{\CC}{\R^N_+}
\newcommand{\FF}{\mathcal F}
\newcommand{\Sc}{\mathcal S}
\newcommand{\PSET}{\mathcal P}

\newcommand{\eps}{\varepsilon}

\newcommand{\diag}{\operatorname{{\mathrm diag}}}

\usepackage{color}

\doublespace

\begin{document}
\title{Entrainment in the Master Equation  
\thanks{This research   is partially supported by   a research grant  from   the
Israel Science Foundation~(ISF grant 410/15)}}

\author{  Michael Margaliot,  Lars Gr\"une and Thomas Kriecherbauer\IEEEcompsocitemizethanks{
\IEEEcompsocthanksitem
M. Margaliot (Corresponding Author)
is with the School of Electrical Engineering and the Sagol School of Neuroscience, Tel-Aviv
University, Tel-Aviv 69978, Israel.
E-mail: \texttt{michaelm@eng.tau.ac.il}
\IEEEcompsocthanksitem
  L. Gr\"une and T. Kriecherbauer are with the Mathematical Institute, University of Bayreuth, 95440 Bayreuth, Germany. E-mail: \texttt{lars.gruene, thomas.kriecherbauer@uni-bayreuth.de}
 }}

\maketitle
\begin{abstract}
The master equation plays an important role in many scientific fields
including physics, chemistry, systems biology, physical finance,  and sociodynamics.
We consider the master equation with periodic transition rates.
This may  represent an external periodic excitation like the 24h solar day in biological systems or
periodic traffic lights in a model of   vehicular traffic. 
Using tools from systems and control theory,
we  prove  that under mild technical conditions every solution
of the master equation converges to a periodic solution with the same period as the rates. In other words, the master equation
entrains (or phase locks) to periodic excitations. We describe two applications of our theoretical results to 
important models from  statistical mechanics and epidemiology. 
 \end{abstract}



\section{Introduction}\label{sec:intro}
 Consider  a physical system that can be in one of exactly~$N$ possible configurations:~$C_1,\dots,C_N$. 
Let~$x_i(t) \in[0,1]$ denote the probability that the system is in configuration~$i$ at time~$t$. 
Let~$x(t):=\begin{bmatrix} x_1(t)&\dots& x_N(t) \end{bmatrix}'$ denote
 the   (column) state-vector of probabilities at time~$t$.

The \emph{master equation}  describes the time evolution  of these probabilities:
\begin{align}\label{eq:meq}
										\dot x_1(t)&=  
										\sum_{ \mycom{j =1 }{j  \not  = 1 }}^N  p_{j1}(t,x(t))x_j(t)
										-\sum_{\mycom{j   =  1 }{j  \not  = 1} }^N  p_{1j}(t,x(t)) x_1(t)    ,\nonumber  \\
										&\vdots \nonumber\\
										\dot x_N(t)&=  
										\sum_{\mycom{j =1}{  j  \not  = N} }^N  p_{jN}(t,x(t))x_j(t)
										-\sum_{ \mycom{j   =  1 }{ j  \not  = N} }^N  p_{Nj}(t,x(t)) x_N(t)   ,
\end{align}
where~$p_{ji}(t,x(t))\geq 0$ denotes the rate of transition from configuration~$C_j$ to configuration~$C_i$.
We assume a general case where the transition rates at time~$t$ 
  may depend on both~$t$ and  the state~$x(t)$. This makes~\eqref{eq:meq}  a time-varying nonlinear   dynamical system.

	Define~$H:\R^N\to \R$ by~$H(y):=y_1+\dots+y_N$.
Since~$x_i$ represents the probability of being in configuration~$C_i$, we assume that the initial condition~$x(t_0)$
satisfies~$H(x(t_0))= x_1(t_0)+\dots+x_N(t_0)=1$. 
Eq.~\eqref{eq:meq} then implies that
	\be\label{eq:sumzerp}
						\sum_{i=1}^N \dot x_i(t)\equiv 0,
	\ee
so~$H(x(t,t_0,x(t_0)))\equiv 1$, that is,~$H$ is a \emph{first integral} of~\eqref{eq:meq}. 
This simply means  that  summing   the probabilities of being at configuration~$C_i$ over all possible~$i$
yields one.  
	
	The master equation can be explained intuitively   as 
describing the balance of probability
currents going in and out of each possible state.
A rigorous derivation  for   a chemically reacting gas-phase
system that is kept well stirred and in thermal equilibrium is 
given in~\cite{GILLESPIE1992404}.
The master equation plays a fundamental role in 
  physics (where it is sometimes referred to as the Pauli
	master equation), chemistry, systems biology, sociodynamics, and more. See e.g. the  monographs~\cite{meq_book,stokamp} for more details.

In the special  case where~$p_{ij}(t,x) = p_{ij}(t)$ for all $i$, $j$ system \eqref{eq:meq} 
 is related to a Markov process in the following way. Denote by~$P_{\tau}$ the fundamental matrix of \eqref{eq:meq}
	with~$P_{\tau}(\tau) = I_N$. Then
$P_{\tau}(t)$ is a stochastic matrix (i.e. the sum of every column of $P_{\tau}(t)$ is equal to one)  for~$t \geq \tau$. 
The obvious relation~$P_{s}(t) P_{\tau}(s) = P_{\tau}(t)$ for~$\tau < s < t$ encodes  the Chapman-Kolmogorov equations if we interpret $(P_{\tau}(t))_{ij}$ as transition probabilities for a system to be in configuration~$C_i$ at time~$t$, provided it is in configuration~$C_j$ at time~$\tau$. Together with an initial probability distribution on the states~$1, \ldots, N$, the transition probabilities then define a unique Markov process and equation~\eqref{eq:meq} is called its \emph{forward equation}~\cite{adven_stoch,book_mastereq}.

	In many physical  systems the number of possible configurations~$N$ can be very large. 
	For example, the    well-known  
	\emph{totally asymmetric simple exclusion principle}~TASEP model (see, e.g.~\cite{TASEP_book,krug2016}   and the references therein)
 includes a lattice of~$n$ consecutive 
	sites, and each site can be either free or occupied by a particle, so the number of possible 
	configuration is~$N=2^n$. 
	In such cases,   simulating the master equation   and calculating its steady-state may be difficult and  
	special 
	methods must be applied (see, e.g.~\cite{nadler_schulten1986,krug2016}).

	Here, we are interested in deriving  theoretical results  that hold  for any~$N$. 
Specifically, we  consider the case where the transition rates are periodic with a common period~$T > 0$, that is,
\be\label{eq:trperix}
				p_{ij}(t+T,x)=p_{ij}(t,x),
\ee
for all~$i,j$, all~$t $, and all~$x$.  
We refer to~\eqref{eq:meq} with the rates satisfying~\eqref{eq:trperix}
as the \emph{$T$-periodic master equation}.
Note that this  includes
the case where one [or several] of the rates is [are]~$T$-periodic with~$T>0$, 
and the other rates are time-independent, as a time-independent function satisfies~\eqref{eq:trperix} for all~$T$.
Clearly, from \eqref{eq:trperix} it also follows that $p_{ij}(t+kT,x)=p_{ij}(t,x)$ for any integer $k$. In order to make the \emph{period} a well defined notion one therefore often 
 requires that the period is the \emph{minimal} real number $T>0$ for which \eqref{eq:trperix} is satisfied. Then constant functions do not have a period. Since we want to
 include here the case of time-independent transition rates, see e.g.~Corollary~\ref{cor:special_cases} below, we do not require the minimality of the 
 common period~$T$ in~\eqref{eq:trperix}.
 
We consider  the problem of entrainment (or phase-locking) in the~$T$-periodic master equation.
\begin{Problem}\label{prob:rnter} 
Given a system described by a $T$-periodic master equation,
 determine if for every initial condition
the probabilities~$x_i(t)$, $i=1,\dots,N$,  
converge to a periodic solution  with   period~$T$. If this is so, determine if the periodic solution is unique or not. 
\end{Problem} 

In other words, if we view the transition rates as a $T$-periodic excitation then the problem is to determine if the state of the system entrains, that is,  converges to a periodic trajectory with the same period~$T$.  If this is so, an important question is whether there exist a unique periodic trajectory~$\gamma$
 and then
every solution converges to~$\gamma$.

Entrainment is important in many natural and artificial systems.
For example,   organisms are often exposed to periodic excitations like the~24h
 solar day and the periodic cell-cycle division process. Proper functioning often requires accurate  entrainment  of various biological  processes 
 to this excitation~\cite{entrain2011}.  
   For example, 
cardiac arrhythmias  is a heart disorder occurring when every other pulse generated by the 
sinoatrial  node pacemaker is ineffective in driving the ventricular rhythm~\cite{Keith1984}.

Epidemics of infectious diseases often correlate
 with 
seasonal changes and  the required interventions, such as pulse vaccination, may also need to be periodic~\cite{epidemics_2006}. 
In mathematical population models, this means that the so called \emph{transmission parameter}
 is periodic, with a period  of one year, 
and entrainment means that the spread of epidemics converges to a   periodic pattern 
with the same period.    
 As another example, traffic flow is often controlled by periodically-varying traffic lights.
In this context, entrainment means that the traffic flow converges to a periodic pattern with the same period
as the traffic lights. This observation could be useful for the study of the \emph{green wave  phenomenon}~\cite{sync_traffic}.
Another example, from the
 field of power electronics, involves connecting a synchronous generator to the electric grid. 
The periodically-varying voltage in the grid may be interpreted as a periodic excitation to the generator, and proper functioning requires
the generator to entrain to this excitation (see e.g.~\cite{dorfler16} and the references therein).

In the special case of time-invariant rates Problem~\ref{prob:rnter} 
 reduces to determining if every solution converges
 to a steady-state, and whether there exits a unique steady-state. 
Indeed, time-invariant rates are~$T$-periodic for any~$T>0$
and thus entrainment means     
 convergence  to a solution that is~$T$-periodic for any~$T>0$,
i.e. a steady-state.

Since the~$x_i$s represent probabilities,
\be\label{eq:xgitr}
x_i(t) \in[0,1] \text{ for all $i$ and }  \sum_j x_j(t)=1  
\ee
for all~$t\geq t_0$. 
The structure of the master equation guarantees that if~$x(t)$
satisfies~\eqref{eq:xgitr} at time~$t=t_0$ then~\eqref{eq:xgitr} holds for all~$t\geq t_0$
even when the~$x_i$s are not necessarily linked to probabilities. 
Our results below hold of course  in this case as well.
The next example demonstrates such a case. 

\begin{Example}\label{exa:n2}
			An important topic  in sociodynamics is the formation of large cities due to population migration. 
			Ref.~\cite[Chapter~8]{meq_book} considers  a master equation describing the flow of individuals  between~$N$ settlements. 
			The transition rates~$p_{ij}$ in this model represent the probability per time unit 
			that an individual living in settlement~$i$ will migrate to settlement~$j$. 
			A mean-field approximation of this master equation yields a model in the form~\eqref{eq:meq},
			where~$x_i$ represents the average density at settlement~$i$,
			and~$p_{ij}=\exp( ( x_j-x_i)k_{ij})$, with~$k_{ij}>0$.
This models the fact that 			
  the rate of transition from settlement~$i$ to settlement~$j$ increases when the population in
	settlement~$j$ is larger than in~$i$, i.e. the tendency of individuals to migrate  to larger cities. 
			 Note that the rates here are state-dependent, but not time-dependent. However, 
			it is natural to assume that migration decisions  depend on the season. For example,   
			the tendency to migrate to colder cities  may decrease [increase]  
			in the winter [summer]. This can be modeled 
			by 
		 adding time-dependence, say,  changing the scaling 
		parameters~$k_{ij}$ to functions~$k_{ij}(t)$,
		that are periodic with a period of one year. Then the transition rates depend on both state and time, and are periodic. 
\end{Example}

For small values of~$N$ it is sometimes possible to solve the master equation and then
 analyze entrainment directly. 
The next   example demonstrates this.
\begin{Example}\label{exa:goo}
Consider the master equation~\eqref{eq:meq} with~$N=2$ and continuous time- (but not state-) dependent rates, i.e.~$p_{ij}=p_{ij}(t)\geq 0$.
Then~\eqref{eq:meq} can be written as 
\begin{align}\label{eq:myta}
										\begin{bmatrix}\dot {x}_1\\ \dot{x}_2 \end{bmatrix}=
										\begin{bmatrix}  -p_{12} &p_{21} \\ p_{12} & -p_{21} \end{bmatrix}
											\begin{bmatrix}  {x}_1\\ {x}_2 \end{bmatrix}.
\end{align}
Assume also that all the rates are periodic 
with period~$T>0$.
Using the fact that~$x_1(t)+x_2(t)\equiv 1$ yields
\be\label{eq:xsrr}
				\dot x_1(t)= p_{21}(t)-(p_{12}(t)+p_{21}(t))x_1(t).
\ee
Recall that~$x_1(0),x_2(0) \in [0,1]$ with~$x_1(0)+x_2(0)=1$. Eq.~\eqref{eq:xsrr} implies 
that~$x_1(t)\in[0,1] $ for all~$t\geq 0$,
and thus~$x_2(t)\in [0,1]$ for all~$t\geq 0$.
Solving~\eqref{eq:xsrr}
yields
\begin{align}\label{eq:soltyp}
							x_1(t)&=   \exp\left(     -\int_0^t     (p_{12}(s)+ p_{21}(s)) \diff s   \right )  (x_1(0)+c(t)), \\
							x_2(t)&=1-x_1(t),\nonumber 
\end{align}
where~$c(t):= \int_0^t  p_{21}(\tau)    \exp\left (   \int_0^\tau (p_{12}(s)+ p_{21}(s)) \diff s \right )   \diff \tau    $. 
To analyze if every solution converges to a periodic solution we consider    two cases.

\noindent Case 1: If~$p_{12}(t)+p_{21}(t)\equiv 0$  then~\eqref{eq:soltyp} yields~$x(t)\equiv x(0) $, i.e. every point in the state-space is an equilibrium point. 
This means that every solution is a periodic solution with period~$T$. 
 
\noindent Case 2: Assume that there exists a time~$t^*\in[0,T)$ such that~$p_{12}(t^*){+}p_{21}(t^*)> 0$. (Note that 
 by continuity this in fact holds on a time interval that includes~$t^*$). 
 The solution~\eqref{eq:soltyp} is periodic with period~$T$ 
if and only if~$x_1(T)=x_1(0)$ i.e. if and only if 
\be\label{eq:x1t}
x_1(0)=\frac{ \exp\left(     -\int_0^T     (p_{12}(s)+ p_{21}(s)) \diff s   \right )  c(T) 
}{1- \exp\left(     -\int_0^T     (p_{12}(s)+ p_{21}(s)) \diff s   \right )
  }.
\ee
It is straightforward to show that the right-hand side in this equation is in~$[0,1]$, so 
  in this case there exists a unique periodic trajectory~$\gamma(t)$, with~$\gamma_1(0)$ equal to the expression in~\eqref{eq:x1t}, and~$\gamma_2(0)=1-\gamma_1(0)$.  
To determine if every trajectory converges to~$\gamma$, let~$z(t):=x(t)-\gamma(t)$.  Then
\[
					\dot z_1= -(p_{12}+p_{21})z_1,\quad z_1(0)=x_1(0)-\gamma_1(0).
\]
Since~$p_{12}(t){+}p_{21}(t)$ is  positive on a time interval,
 and $T$-periodic,~$z_1(t)$ converges to zero, and we conclude that any trajectory of the system
converges to the unique periodic solution~$\gamma$.
\end{Example}
Of course, when~$N>2$ and the rates depend on both~$t$ and~$x$ this type of explicit analysis is impossible, and the proof of entrainment requires a different approach.

In general, 
proving that a time-varying nonlinear dynamical system  entrains  to periodic excitations is non trivial. 
Rigorous proofs are known  for two classes of dynamical systems: contractive systems, and monotone systems admitting a  first integral.

A system is called \emph{contractive} if any two trajectories approach one another at an exponential 
rate~\cite{sontag_cotraction_tutorial,LOHMILLER1998683}. Such systems entrain to
 periodic excitations~\cite{entrain2011,coogan_margaliot}.
An important special  case is asymptotically stable linear systems with an additive periodic input~$u$, that is, systems in the form
\be\label{eq:linss}
\dot x=Ax+B u,
\ee
with~$x\in \R^N$, $A\in\R^{N\times N}$ a Hurwitz matrix,\footnote{i.e. the real part of every eigenvalue of~$A$ is negative.}
 $u\in\R^M$, and~$B\in \R^{N\times M}$.
In this case~$x(t)$ converges to a periodic solution~$\gamma(t)$ and it is also
  possible to obtain a closed-form description of~$\gamma$ using the transfer function of the linear system~\cite{zadeh_book_1963}. 
  We note that even in the case that the $p_{ij}$s in~\eqref{eq:meq} do not depend on~$x$, i.e., when~\eqref{eq:meq} is linear in~$x$, the master equation is not of the form~\eqref{eq:linss}
	because the periodic influence in~\eqref{eq:meq} enters through    the transition rates
	 $p_{ij}$ and  not through  an additive input channel.

 A system is called 
\emph{monotone} if its flow preserves  
 a partial order, induced by an appropriate cone~$K$, between its initial conditions~\cite{hlsmith}. 
An important special case are cooperative systems for which the cone~$K$ is the positive orthant. 
 Cooperative systems that admit a first integral entrain to periodic excitations. 
It is interesting to note
that   proofs of this property often follow from contraction arguments~\cite{Mierc1991}.

 The master equation~\eqref{eq:meq} is in general not contractive,
although as we will show in Theorem~\ref{thm:T3}  below  it is on the ``verge of contraction'' with respect to the~$\ell_1$ vector norm (see~\cite{cast_book} for some related considerations).
However,~\eqref{eq:meq} admits a first integral and is often a cooperative system (see Theorem~\ref{thm:T4} below). In  
particular, when the rates do not depend on the state, i.e.~$p_{ij}= p_{ij}(t)$  then~\eqref{eq:meq} is always cooperative.

Although  entrainment has attracted enormous research attention, it seems that it has not been addressed before for the general case of 
  systems     modeled using a~$T$-periodic 
 master equation. Here we apply the theory of cooperative  dynamical systems admitting a first integral
to derive conditions guaranteeing that 
  the answer   to    Problem~\ref{prob:rnter} is affirmative.
In Section~\ref{sec:app}, we describe   two  applications of our approach to   important
 systems from statistical physics. 
The first is    the \emph{totally asymmetric simple exclusion process}~(TASEP).
This model has been introduced in the context of bio-cellular processes~\cite{MacDonald1968},
and has become the standard model for  the flow  of ribosomes
along the mRNA molecule during translation~\cite{TASEP_tutorial_2011,Shaw2003}. More generally, TASEP has become 
 a paradigmatic model for the statistical mechanics of nonequilibrium systems~\cite{TASEP_book, krug2016, kriecherbauer_krug2010}.
 It is in particular used to study the stochastic dynamics of interacting particle 
systems such as vehicular traffic~\cite{traffic_TASEP_99}.

The second application is to an important model from epidemiology called the
  stochastic susceptible-infected-susceptible~(SIS) model.

The remainder of this paper is organized as follows. 
In the next section we state the exact mathematical formulation of the
master equation that we assume throughout together with our main results Theorems~\ref{thm:mptra}
 and~\ref{thm:irred}. Section~\ref{sec:app} 
 describes the two applications to statistical physics and to epidemiology   mentioned above. 
This  is followed by a brief discussion of the significance of the  results and an outlook on possible future directions of research. 
The Appendix includes all the proofs.
These are based on known tools, 
  yet we are able to use the special  structure of the master equation to derive stronger results than those available in the literature on monotone dynamical systems. 
\section{Main results}\label{sec:main}

We begin by specifying the exact conditions on~\eqref{eq:meq} that are assumed throughout. 
For any time~$t$, $x(t)$ is an~$N$-dimensional column  vector that includes
the probabilities of all~$N$ possible configurations.
The relevant state-space is thus
\[
			\Omega:=\{ y\in \R^N   \,|\,   y_i \geq 0 \text{ for all } i, \text{ and }  \sum_{i=1}^N y_i=1 \}.
\]
For an initial time~$t_0\geq 0 $ and an initial condition~$x(t_0)$, let~$x(t;t_0,x(t_0))$
denote the solution of~\eqref{eq:meq} at time~$t\geq t_0$. 
For our purposes  it will be convenient to assume
that the vector field associated with system~\eqref{eq:meq} is not only defined on the set~$\Omega$,
but on all the closed positive cone
\[ \CC := \{ x\in\R^N  \,|\,   x_j\ge 0 \mbox{ for all } 1\le j\le N\}\,. \]

Throughout this paper we assume that the following condition holds.
\begin{Assumption}\label{assump:tou}
There exists~$T > 0$ such that 
  the transition rates~$p_{ij}(t,{x})$ are:
	continuous and non-negative on~$[0,T) \times \R^N_+  $;
		continuously differentiable with respect to~${x}$
	on~$[0,T) \times \interior (\R^N_+ ) $ and the derivative admits  a continuous extension onto $[0,T) \times \CC$; 
	and  are jointly   periodic with period~$T $, that is, 
\be\label{eq:trperi}
				p_{ij}(t+T,{x})=p_{ij}(t,{x}),
\ee
for all~$i,j$, all~$t\in [0,T)$, and all~${x}\in \R^N_+$.  
\end{Assumption}

  Let~$\inter \Omega$ denote the relative
interior of~$\Omega$,
that is,
\[
		\inter \Omega=\{ y\in \R^N  \,|\, y_i> 0 \text{ for all } i, \text{ and }  \sum_{i=1}^N y_i=1 \}.
\]
Note that if  the rates are only   defined  on~${x}\in \Omega$, with
partial derivatives with respect to ${x}_j$     on $\inter(\Omega$) with continuous extensions to $\Omega$,
then they can be extended to~$\R^N_+$ so that the conditions in Assumption~\ref{assump:tou} hold. For example,
by defining them to be
 constant on rays through the origin and multiplied by a cut-off function $\chi(|{x}|_1)$, where   $\chi$ is a smooth 
function with a compact support in~$[0, \infty)$, satisfying  $\chi(s)=1$ for~$s=1$, and where $|x|_1$ denotes the $\ell_1$-norm of $x$.

We now determine the conditions guaranteeing that~\eqref{eq:meq}  is a cooperative dynamical system.
Note  that~\eqref{eq:meq} can be written as
	\be\label{eq:memlin}
	\dot x(t)=f(t,x):=A(t,x(t))x(t),
	\ee
	where~$A \in \R^{N\times N}$ is a matrix with entries
\be\label{eq:aij}
			a_{ij}(t,x(t)):=\begin{cases} 
							p_{ji}(t,x(t)),&			\text{ if } i\not =j, \\
							\displaystyle -\sum_{ \mycom{k=1 }{ k \not = i }}^N p_{ik}(t,x(t)),&			\text{ if } i =j.  
			\end{cases}
\ee
The Jacobian   of the vector-field~$f$ is the $N\times N$ matrix 
\be\label{eq:jx}
			J(t,x):=\frac{\partial f(t,x)}{\partial x}= A(t,x) + B(t,x),
\ee
where~$B $ is the matrix with entries~$b_{ij} :=  \sum_{k=1}^N  x_k   \frac{\partial a_{ik}}{\partial x_j} $.
Recall that a matrix~$M\in\R^{n\times n}$ is called \emph{Metzler}
if every off-diagonal entry of~$M$ is non-negative. 
It follows that
if~$p_{ji}(t,x)+\sum_{k=1}^N x_k \frac{\partial a_{ik}(t,x)}{\partial x_j} \geq 0  $
for all~$i\not =j$, all~$t\geq t_0$ and all~$x\in\Omega$ then~$J(t,x)$ is Metzler   for all~$t\geq t_0$ and all~$x\in\Omega$.

We can now state our first result. 
\begin{Theorem}\label{thm:mptra}
Suppose that 
\be\label{eq:condper}
p_{ji}(t,x)+\sum_{k=1}^N \frac{\partial a_{ik}(t,x)}{\partial x_j}x_k\geq 0  \text{ for  all } i\not=j , \;t\geq t_0 , \; x\in\Omega .
\ee
Then for any~$t_0\geq  0 $ and any~$x(t_0)\in \Omega$ the solution~$x(t;t_0,x(t_0))$ of~\eqref{eq:meq}
converges to a periodic solution with period~$T$.
\end{Theorem}

If the rates depend on time, but not  on the state, i.e.~$p_{ij}=p_{ij}(t)$ for all~$i,j$, then 
the condition in Thm.~\ref{thm:mptra} always   holds, and this yields the 
following result.

\begin{Corollary}\label{cor:special_cases}
If~$p_{ij}=p_{ij}(t)$ for all~$i,j$ then
for any~$t_0\geq  0 $ and any~$x(t_0)\in \Omega$ the solution~$x(t;t_0,x(t_0))$ of~\eqref{eq:meq}
converges to a periodic solution with period~$T$.
\end{Corollary}

If  the rates depend on the state, but not on time then 
we may apply Thm.~\ref{thm:mptra} for all~$T > 0$. 
Thus, the trajectories converge to a periodic solution with an arbitrary period, i.e. a 
steady-state. 
 This yields the 
following result.

\begin{Corollary}\label{cor:special_cases2}
If~$p_{ij}=p_{ij}(x)$ for all~$i,j$ and   in addition condition \eqref{eq:condper} holds
 then
for any~$t_0\geq  0 $ and any~$x(t_0)\in \Omega$ the solution~$x(t;t_0,x(t_0))$ of~\eqref{eq:meq}
converges to a steady state.
\end{Corollary}

In some applications, it is useful to establish that
   all trajectories of~\eqref{eq:meq}
	converge to a \emph{unique} periodic trajectory. Recall that 
a matrix $M\in \R^{n\times n}$, with~$n\geq 2$,  is said to be
\emph{reducible} if there exist a permutation matrix $P\in \{0,1\}^{n\times n}$,
and an integer  $1\leq r\leq n-1$ such that $P' MP=\begin{bmatrix}
B & C \\
0 & D%
\end{bmatrix}$,
where $B\in \R^{r\times r}$, $D\in \R^{(n-r)\times
(n-r)}$, $C\in \R^{r\times (n-r)}$, and $0\in \R^{(n-r)\times r}$
is a zero matrix.
A matrix   is called \emph{irreducible} if
it is not reducible.
It is well-known that a Metzler matrix~$M$ is irreducible if and only if   
the  graph associated with the adjacency matrix of $M$  is strongly connected,
see \cite[Theorems 6.2.14 and 6.2.24]{mat_ana_sec_ed}.

\begin{Theorem}\label{thm:irred} 
Suppose that the conditions in Theorem~\ref{thm:mptra} hold
and, furthermore,  that there exists a time~$t^*\geq t_0$
such that~$A(t^*,{x})+B(t^*,{x})$ is an irreducible matrix for  all~${x} \in \Omega$. 
Then~\eqref{eq:meq} admits a unique  periodic solution~$\gamma $ in~$ \Omega$, with period~$T$,
and every
  solution~$x(t;t_0,x(t_0))$ with~$x(t_0)  \in \Omega$
converges to~$\gamma$ at an exponential rate. 
\end{Theorem}

\begin{Example}
Consider the system in Example~\ref{exa:goo}.
This is of the form~\eqref{eq:memlin} with
\[
			A(t)=\begin{bmatrix}  -p_{12}(t) &p_{21}(t) \\ p_{12} (t)& -p_{21} (t)\end{bmatrix}.
\]
If there exists a time~$t^*$ such that~$p_{12}(t^*), p_{21}(t^*) >0$ then~$A(t^*)$
 is irreducible. We conclude that in this case all the conditions in Theorem~\ref{thm:irred} 
hold, so the system admits a unique $T$-periodic solution~$\gamma$
 and every trajectory converges to~$\gamma$.   This agrees of course with the results of the analysis in Example~\ref{exa:goo} above 
where we arrived at the same conclusion under the slightly weaker assumption that~$p_{12}(t^*)+p_{21}(t^*) >0$. 
\end{Example}

 The next section describes   applications of our results to two important
models. 
\section{Applications}\label{sec:app}

\subsection{Entrainment in TASEP}\label{ssec:app.1}


The \emph{totally asymmetric simple exclusion process}~(TASEP)
 is a stochastic model for particles hopping along a 1D chain.
A particle at site $s_k$ hops
to site $s_{k+1}$ (the next site on the right)
  with an exponentially distributed probability\footnote{We consider the continuous time version of TASEP here.} with rate $h_k$, provided the site $s_{k+1}$ is not occupied by another particle. 
	This simple exclusion property generates an indirect link between the particles and allows
	to model the formation of traffic jams. Indeed, 
	if a particle ``gets stuck'' for a long time in the same site then other    particles accumulate behind it.

	At the left end of the chain  particles enter with a certain entry rate~$\alpha>0$ and at the right end particles leave with a rate~$\beta>0$. 
As pointed out in the introduction, TASEP has become a standard model for modeling ribosome  flow during translation, and is a paradigmatic model for the statistical mechanics of nonequilibrium systems. 
 We note that in the classical TASEP model the rates~$\alpha$, $\beta$, and~$h_i$ are constants,
but several papers considered
 TASEP with periodic rates~\cite{PhysRevE.78.011122,PhysRevE.93.012123,PhysRevE.83.031115}
 that can e.g. be used as models for vehicular  traffic 
controlled by periodically-varying  traffic signals. 

It was shown in~\cite{RFM_entrain}
 that the  dynamic mean-field approximation of   TASEP, called the \emph{ribosome flow model}~(RFM),
 entrains. However, the~RFM  is not a master equation and the proof of entrainment in~\cite{RFM_entrain}
is based on different ideas. For more on the analysis of the~RFM, see e.g.~\cite{RFM_concave,RFM_model_compete_J,RFM_sense,10.1371/journal.pone.0166481}.

For a chain of length $n$, denoting an occupied site by $1$ and a free site by $0$, the set of possible configurations
 is $\{0,1\}^n$,  and thus the number of possible configurations is $N=2^n$. 
The dynamics of TASEP can be expressed as a master equation with transition 
rates~$p_{ij}$ that depend on the values $\alpha$, $\beta$ and~$h_i$, $i=1,\ldots,n$. For the sake of simplicity, we will show this in the specific case~$n=2$,
 but all our
results below hold for any value of~$n$.

When~$n =2$ the possible  configurations of particles along the chain
are~$C_1:=(0,0)$, $C_2:=(0,1)$, $C_3 := (1,0)$ and $C_4 := (1,1)$. 
Let~$x_i(t)$ denote the probability that the system is in
   configuration~$C_i$ at time~$t$, for example,
$x_1$ is the probability that both sites are empty. Then~$x_1$
   may decrease [increase] due to the transition~$C_1 \to C_3$ [$C_2 \to C_1$], i.e. 
when a particle enters the first site [a 
	particle in the second site hops out of the chain].
	This gives
\[
\dot x_1(t) =  - \alpha x_1(t)+\beta x_2(t) .
\]

 Similar considerations for all configurations   lead to the master equation  $\dot x=A x$, with
\[
 A := \left( \begin{array}{cccc} 
-\alpha  & \beta & 0 & 0\\
0 & -\alpha-\beta & h_1 & 0 \\
\alpha & 0 & -h_1 & \beta\\
0 & \alpha & 0 & -\beta
\end{array}\right). 
\]
If the entry, exit and hopping rates are time-dependent and periodic, all with the same period~$T$, one easily sees that the resulting master equation satisfies Assumption~\ref{assump:tou} as well as all assumptions of Theorem~\ref{thm:mptra}. Hence, we   conclude that every solution of the master equation starting in $ \Omega$ converges to a periodic solution with period~$T$. Moreover, if there exists a time~$t^*$
such that~$\alpha(t^*),\beta(t^*),h_1(t^*)>0$ then~$A(t^*)$ is irreducible.
Hence, the conditions of Theorem~\ref{thm:irred} are also satisfied, so 
we   conclude that the periodic solution is unique and convergence takes place at an exponential rate.
It is not difficult to show that the same holds for~TASEP with any length~$n$.

\begin{figure}[t]
 \begin{center}
 \includegraphics[scale=0.75]{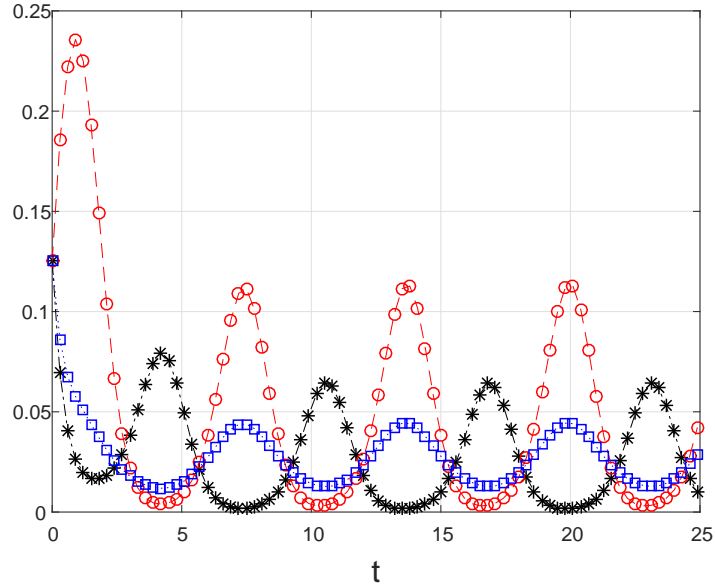}
\caption{Probabilities  $x_1(t)$ (black asterisk‬‏), $x_4(t)$ (blue square), and~$x_8(t)$ (red circle) as a function of~$t$
in Example~\ref{exa:ttad}.}\label{fig:tasep}
\end{center}
\end{figure}

\begin{Example}\label{exa:ttad}
When~$n =3$ the possible      particle configurations
are~$C_1:=(0,0,0)$, $C_2:=(0,0,1)$, $C_3 := (0,1,0)$, $C_4 := (0,1,1)$, $\dots$, $C_8:=(1,1,1)$. 
Let~$x_i(t)$ denote the probability that the system is in
   configuration~$C_i$ at time~$t$. The TASEP master equation in this case is~$\dot x=A x$, with
	\[
	A=\begin{bmatrix}
									-\alpha& \beta& 0 & 0 & 0 & 0 & 0 & 0 \\
    0 & -\alpha-\beta& h_2&  0 & 0 & 0 & 0 & 0 \\
    0 & 0 & -\alpha-h_2& \beta& h_1& 0 & 0 & 0 \\
    0 & 0 & 0 & -\alpha-\beta& 0 & h_1& 0 & 0 \\
    \alpha & 0 & 0 & 0 & -h_1& \beta& 0 & 0 \\
    0 & \alpha& 0 & 0 & 0 & -h_1   -\beta & h_2 &0 \\
    0 & 0 & \alpha& 0 & 0 & 0 & -h_2 &\beta\\
    0 & 0 & 0 & \alpha& 0 &  0 & 0 & -\beta
	\end{bmatrix}.
	\]
	We simulated this system with the rates
		\[
		\alpha(t)=1
		+\cos(t),\; \beta(t)=1+\cos(t+\pi),\;  h_1=1/2, \;  h_2=1/4,
		\]
		and initial condition~$x(0)=\begin{bmatrix} 1/8&\dots&1/8\end{bmatrix}'$. Note that all the rates here are jointly periodic with 
		period~$2\pi$.
Fig.~\ref{fig:tasep} depicts $x_1(t)$ (black asterisk‬‏), $x_4(t)$ (blue square),
 and $x_8(t)$ (red circle) as a function of~$t$ (we depict only three~$x_i$s to avoid cluttering the figure).
 Note that since the entry rate~$\alpha(t)$ is maximal 
and the exit rate~$\beta(t)$ is  minimal at~$t=0$, the probability~$x_8(t)$ [$x_1(t)$] to be in
 state~$(1,1,1)$ [$(0,0,0)$] quickly increases [decreases] near~$t=0$.  As time progresses,
  the probabilities converge to a periodic pattern with period~$2\pi$.
\end{Example}

Entrainment of the probabilities~$x_i$
 has consequences for  other quantities of interest in statistical mechanics. For instance, an imporatnt 
quantity is  the occupation density, i.e., the probability that site $s_k$ is occupied, often denoted by $\langle \tau_k \rangle$, cf.\ \cite{solvers_guide,DerE97}. Denoting the $k$-th component of the configuration $C_i \in\{0,1\}^n$ by $C_{i,k}$, a straightforward computation reveals that
\[ \langle \tau_k(t) \rangle  = \sum_{i=1}^{N} C_{i,k} x_i(t).\] 
It is thus immediate  that     the occupation densities also converge to a unique periodic solution.

This phenomenon has already been observed empirically    in
Ref.~\cite{PhysRevE.78.011122}  that studied a  
  semi-infinite and finite
	TASEP   coupled at
the  end  to  a  reservoir  with  a periodic time-varying  particle  density. This models for example 
a traffic lane ending with   a  periodically-varying traffic light. The simulations 
in~\cite{PhysRevE.78.011122}   suggest  that this leads to the development of a  
sawteeth density profile along the chain, and that
``The  sawteeth  profile  is
changing  with  time,  but  it  regains  its  shape  after  each complete period\ldots''~\cite[p.~011122-2]{PhysRevE.78.011122} 
(see also~\cite{PhysRevE.93.012123,PhysRevE.83.031115} for some related considerations).

Our  results can also be interpreted in terms of  the particles 
along the chain in TASEP. Since the expectation of the occupation densities $\langle \tau_k \rangle$ converges to a periodic solution, this means that in the long term the TASEP dynamics  ``fluctuates'' around a periodic ``mean'' solution (see e.g. the simulation results depicted in  Figure~5 in \cite{RFM_entrain}).
 Moreover, in~\cite{PhysRevE.93.012123,PhysRevE.83.031115}
 it was found for closely related models that the limiting periodic density profiles (whose 
 existence is also guaranteed by our results)
have an interesting structure that   depends  in a non-trivial way on the
  frequency of the transition rates.

\subsection{Entrainment in a stochastic SIS model} 

The stochastic susceptible-infected-susceptible~(SIS) model 
plays an important role in mathematical epidemiology~\cite{nasell}. 
But, as noted in~\cite{Baca2015}, it is usually studied under the assumption of fixed contact and recovery rates.
Here, we apply our results to prove   entrainment in an~SIS model with periodic  rates.

Consider a population of size~$N $ divided into susceptible and infected individuals. 
Let~$S(t)$ [$I(t)$] denote the
 size of the susceptible [infected] part of the population   at time~$t$, so that
$S(t)+I(t)\equiv  N$.
We assume two mechanisms for infection. The first is by contact with an infected and depends on the  contact rate~$a(t)$. 
The second is by some external agent (modeling, say, insect bite) with rate~$c(t)$. 
The recovery rate is~$b(t)$. 
We assume that~$a(t)$, $ b(t)$, and~$c(t)$  are  continuous 
and take non-negative values for all time~$t$.

 If $I (t) = n$ (so~$S(t)=N-n $) then the probability that one individual  recovers in the  time
interval 
$[t , t + \diff t]$
 is $b(t) n \diff t + o(\diff t)$, and  the probability for one new infection to occur in this time interval
is~$a(t) n \frac{ N-n} {N}   \diff t +c(t)    \frac{ N-n} {N}   \diff t+ o(\diff t )$.  
For~$n\in \{ 0,\dots,N\}$, let~$P_n(t)$ denote the probability that~$I(t)=n$.
 This yields the master equation:
\be\label{eq:mased}
					\dot {P}_n=(  (n-1)a +c)\left (1-\frac{n-1}{N}  \right )P_{n-1}-
					\left ( ( n  a+c) (1-\frac{n}{N})+ n b \right ) P_{n} +  (n+1) b P_{n+1},
\ee
for~$n\in\{0,1,\dots , N\} $,  where we define~$P_{-1}=P_{N+1}:=0$, and for simplicity omit the dependence on~$t$. 
This set of~$N+1$ equations may be written in matrix form as
\[
			\dot x=M x  ,
\]
where~$x :=\begin{bmatrix} P_0&P_1&  \dots&P_N \end{bmatrix} '\in [0,1]^{N+1}$, 
$M:=P-D$, with~$D:=\diag(0,b,2b,\dots,Nb)$, 
and~$P $ is the~$(N+1)\times(N+1)$ matrix:
\[
		 \begin{bmatrix} 
					-cq_0 & b & 0 & 0  &\dots & 0 & 0 &0 \\	
		      cq_0 & -  (a+c)q_1 & 2b & 0  &\dots & 0 & 0 &0 \\	
					0 &   (a+c)q_1 & -(2a+c)q_2  & 3b  &\dots & 0 & 0 &0 \\	
					0 & 0  & (2a+c)q_2  &-(3a+c)q_3    &\dots & 0 & 0 &0 \\	
			&&&&\vdots\\
					0 & 0  & 0 & 0   &\dots & ((N-2)a+c) ) q_{N-2}    &   -((N-1) a+c ) q_{N-1}        &bN \\	
					0 & 0  & 0 & 0   &\dots &  0&   ( (N-1)a+c) q_{N-1}    &0	
			\end{bmatrix},
\]
where~$q_i:=1-\frac{i}{N}$. Note that~$M(t)$ is  Metzler, as~$a(t),b(t)$ and~$c(t)$ are non-negative for all~$t$.
Thus, Theorems~\ref{thm:mptra} and~\ref{thm:irred}  yield the following result. 
\begin{Corollary}
If~$a(t)$, $b(t)$ and~$c(t)$ are all~$T$-periodic then any 
solution of~\eqref{eq:mased}   with  $x(0)\in \Omega \subset \mathbb{R}^{N+1}$
  converges to a~$T$-periodic solution. Furthermore, if there exists a time~$t^*\geq 0 $ such that
	\be \label {eq:irff}
	b(t^*)c(t^*)>0
	\ee
	then
	there exists a unique  $T$-periodic solution~$\gamma $  in~$\Omega$ and every solution converges to~$\gamma$. 
\end{Corollary}
 
\begin{figure}[t]
 \begin{center}
 \includegraphics[scale=0.75]{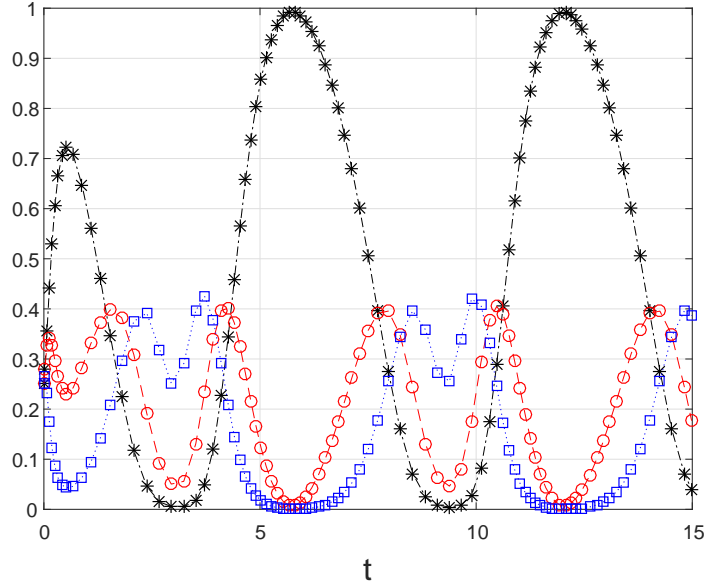}
\caption{Probabilities  $P_0(t)$ (black asterisk‬‏), $P_1(t)$ (red circle), and~$P_2(t)$ (blue square) as a function of~$t$
in Example~\ref{exa:sis}.}\label{fig:sis}
\end{center}
\end{figure}

\begin{Example}\label{exa:sis}
Consider the stochastic SIS model with~$N=3$, $a(t)=1$, $b(t)=3+3\cos(t+0.5)$ and~$c(t)=2-2\sin(t+0.75)$. 
These rates are non-negative and jointly~$T$-periodic for~$T=2\pi$ and  clearly there exists~$t^*\geq 0$ such   that~$b(t^*)c(t^*)>0$. 
Fig.~\ref{fig:sis} depicts~$P_i(t)$, $i=0,1,2$, (note that~$P_3(t)=1-P_0(t)-P_1(t)-P_2(t))$ as a function of
time~$t$  for  the initial condition~$P(0)=(1/4)1_4 \in\Omega $.  It may be seen that
every  $P_i(t)$   converges to a   periodic solution with period~$2\pi $.
 Taking  other initial conditions~$x(0)\in\Omega$ 
  yields convergence  to the
same periodic solution. 
\end{Example}

Note that if the irreducibility condition~\eqref{eq:irff}
 does not hold then the system may have several   periodic solutions. 
To see this, consider for example the case~$b(t)=c(t)\equiv 0$.
 Let~$e^i\in\R^{N+1
}$ denote the vector with   entry~$i$ equal to one and all other entries zero. 
Then    both~$x(t)\equiv e^1$ and~$x(t)\equiv e^{N+1}$
are (periodic) solutions of the dynamics.

\section{Discussion}

In his 1929 paper on periodicity in disease prevalence, H. E. Soper~\cite{Soper1929} states: ``Perhaps no events of human experience interest us so continuously,
from generation to generation, as those which are, or seem to be,
periodic''.  Soper also raised the question of whether the observed 
periodicity in epidemic outbreaks is the result of
 a ``seasonal change in perturbing influences, such as might be brought about by school break-up and reassembling, or other annual recurrences?''
In modern terms,   
this amounts to asking whether   the solutions of the   system describing the 
dynamics of the epidemics  entrain to
periodic variations in the  
  transmission parameters.

Here, we studied entrainment  for dynamical systems described by a master equation.
We considered a rather general formulation where the transition rates may depend on both time and state. 
Also, we did not assume any symmetry conditions (e.g. detailed balance conditions~\cite[Ch.~V]{stokamp}) on the rates. 
We note that this formulation implies similar results for non-linear systems as well. Indeed, consider
the time-varying non-linear system:
\be\label{eq:nolinr}
\dot x=f(t,x),
\ee
and assume that~$f(t,0)=0$ for all~$t$. 
Let~$J(t,x):=\frac{\partial f}{\partial x} (t,x)$ denote the Jacobian of the vector field. 
Then
\begin{align*}
				\dot x & = \int_0^1 \frac{d}{ds} f(t,sx) \diff s  \\
				       &= A(t,x) x,
\end{align*} 
where~$A(t,x):=  \int_0^1 J(t,sx)\diff s $. If~$A(t,x)$ has the form~\eqref{eq:aij} then 
the results above can be applied to~\eqref{eq:nolinr}. 

We  proved
that entrainment indeed holds under quite mild technical conditions. This follows from the fact that
the master equation is a cooperative dynamical system admitting a first integral.  
Due to the prevalence of the master equation as a model for natural and artificial phenomena, 
 we believe that 
this result will find many applications. To demonstrate this,
we described  two  applications of our results: a proof of entrainment in TASEP and in a stochastic SIS model.

The rigorous proof that the solutions of the master equation  entrain is 
of course a necessary first step in studying the structure  of the periodic trajectory (or trajectories), and its dependence 
on various parameters.
Indeed, in  many applications  it is 
of interest to obtain more information  on the  periodic trajectory
  e.g. its  amplitude.  Of course, one cannot expect in general
to obtain a closed-form description of the limit cycle. However, for contractive dynamical systems there do exist efficient methods for 
obtaining a closed-form approximation of the limit cycle accompanied by explicit error bounds~\cite{coogan_margaliot}. 
Developing a similar approach for the attractive limit cycle  of the master equation may  be an interesting topic for further research. 
 In the specific case of TASEP with fixed rates, there exists a powerful
representation of the steady-state in terms of a product of  matrices~\cite{solvers_guide,krug2016}.
It may be of interest to try and represent the periodic steady-state using a similar product, but with matrices with periodic entries.
This could be used in particular to study the effects of periodic perturbations to the boundary-induced phase transitions that have been
observed for TASEP in~\cite{krug1991}. 

 \section*{Acknowledgments}
We thank Yoram Zarai for helpful comments.
 The second and third authors are grateful to Joachim Krug for very helpful discussions on interacting particle systems and
for pointing out a number of references.  

\section*{Competing Interests}
The authors have no competing interests.

\section*{Author  Contributions}
MM, LG, and TK performed the research  and wrote the paper.

\section{Appendix: Proofs of Theorems \ref{thm:mptra} and \ref{thm:irred}}
The proofs of Theorems~\ref{thm:mptra} and~\ref{thm:irred} are based on known tools from the theory 
of monotone dynamical systems
admitting a first integral with a positive gradient (see, e.g.~\cite{peri_gross_subs,Dancer1991,mono_periodic_96}). We present in this appendix a self-contained proof   taking full advantage of the technical simplifications that our specific
setting permits. This, in particular, allows us to prove that the results hold on the \emph{closed} state-space and also
that irreducibility at a single time point is enough to guarantee convergence to a unique periodic solution. 
Without loss of generality we always assume that the initial time is~$t_0=0$.
It is convenient to work with the $\ell_1$ vector norm $|{x} |_1=\sum_i |{x}_i|$.

We begin by introducing some notation. First recall   the notation
\[ \CC := \{ x\in\R^N  \,|\,   x_j\ge 0 \mbox{ for all } 1\le j\le N\} \]
for the closed positive cone. Define a  set of vector fields by:
\begin{eqnarray*} \FF & := & \{ f:[0,\infty) \times \CC \to \R^N\,|\, \mbox{properties (i)--(v) below  are satisfied}\}, 
\end{eqnarray*}

where
\begin{enumerate}
\item[(i)] $f$ is continuous;
\item[(ii)] for all $j\in\{1,\ldots,N\}$, $\frac{\partial f}{\partial x_j}(t,x)$ exists for $(t,x)\in[0,\infty)\times \interior (\CC)$ and has a continuous extension onto $[0,\infty) \times \CC$. Thus, $J(t,x)$ from \eqref{eq:jx} is defined on $[0,\infty) \times \CC$ to be the continuous extension of $\frac{\partial f}{\partial x_j}(t,x)$;
\item[(iii)] $J(t,x)$ is Metzler for all $(t,x)\in [0,\infty) \times \CC$;
\item[(iv)] $\sum_{i=1}^N f_i(t,x) = 0$ for all $(t,x)\in [0,\infty) \times \CC$;
\item[(v)] $f(t,0)=0$ for all $t\in [0,\infty)$.
\end{enumerate}
For~$T>0$,  let
$\FF_T   :=   \{f\in\FF  \,|\,   f(t+T,x) = f(t,x) \mbox{ for all } (t,x)\in [0,\infty) \times \CC \}$, that is, the set
of vector fields in~$\FF$ that are also $T$-periodic.

It is straightforward to check that $f(t,x)=A(t,x)x$ with $A$ defined by \eqref{eq:aij} belongs to $\FF_T$ if Assumption \ref{assump:tou}
and the assumptions of Theorem~\ref{thm:mptra} hold. Therefore, Theorem \ref{thm:mptra} follows from the following result.

\begin{Theorem}\label{thm:T1}
 If $f\in\FF_T$ then for all $x_0\in\Omega$ the solution $x(t;x_0)$ of the initial value problem
\[ \dot x = f(t,x), \quad x(0)=x_0 ,
\]
is asymptotically $T$-periodic, i.e., there exists a solution $\gamma:\R\to\Omega$ of $\dot \gamma = f(t,\gamma)$,
 with $\gamma(t+T)=\gamma(t)$ for all $t\in\R$, and 
\[ 
\lim_{t\to\infty} |x(t;x_0) - \gamma(t)|_{{1}} = 0  .
\]
\end{Theorem}

Let
\[ \FF_{\text{irr}}^\Omega := \{ f\in\FF\,|\, \text{there exists } t^*\ge 0 \text{ such that }  J(t^*,x) \text{ is irreducible for all }  x\in\Omega\}.
 \]
The next result is a generalization  of Theorem \ref{thm:irred}.
\begin{Theorem}\label{thm:T2}  
If $f\in (\FF_T\cap \FF_{\text{irr}}^\Omega)$ then
the differential equation $\dot x=f(t,x)$ admits
 a \emph{unique} $T$-periodic solution $\gamma:\R\to\Omega$. Moreover, there exists~$\alpha >0$ such that for any initial condition~$x_0\in\Omega$ the corresponding solution~$x(t;x_0)$ 
satisfies 
\[ 
  |x(t;x_0)-\gamma(t)|_1   < 2 e^{-\alpha t}   |x_0-\gamma(0)|_1 ,
\]
i.e. the solution converges to~$\gamma$ with exponential rate~$\alpha$.
\end{Theorem}


Complete proofs of Theorem \ref{thm:T1} and Theorem \ref{thm:T2} are provided in the following seven subsections. We begin by showing
in Lemma \ref{lemma:T1} that solutions of $\dot x = f(t,x)$, $f \in \mathcal{F}$, that start in the closed state space $\mathbb{R}_{+}^{N}$ are unique
and remain in~$\mathbb{R}_{+}^{N}$ for all positive times. In the second subsection,
 we prove that  for the subset of linear vector fields $f$ in $\mathcal{F}$   the
flow is cooperative, and non-expansive or even contractive in the case of irreducibility. 
The latter property is then generalized to the nonlinear setting (Theorem~\ref{thm:T3}),
 which is enough to prove Theorem~\ref{thm:T2} in Subsection \ref{ssec:appendix.4}. The cooperative behavior for nonlinear vector fields
is stated in Theorem \ref{thm:T4}. In Subsection~\ref{ssec:appendix.6}, we argue that the non-expansiveness of the flow together with 
the existence of a fixed point in the $\omega$-limit set of the period map
implies the asymptotic periodicity of the solution. The proof that such a fixed point exists is deferred to the final subsection.
It uses the cooperative behavior of the flow as well as the fact 
that the first integral $H$ has a positive gradient, i.e.~$\nabla H \in$ int$(\mathbb{R}_{+}^{N})$.

\subsection{Positive invariance of $\CC$}\label{ssec:appendix.1}
Our first goal is to establish in Lemma \ref{lemma:T1}  below
 that for any~$x_0\in\CC$ a unique  solution $x(t;x_0)$ exists for all $t\in[0,\infty)$  and
 remains in the closed cone $\CC$. Denote 
\[ \BB := \left\{ B\in\R^{N\times N}\,\left|\, B \mbox{ is Metzler and } \sum_{i=1}^N B_{ij} = 0 \mbox{ for all } j\in\{1,\ldots,N\} \right.\right\}.\]
\begin{Proposition}\label{prop:T1} 
Assume that~$f\in\FF$. Then there exists a continuous map~$B:[0,\infty) \times\CC\times\CC \to \BB$ such that
\be\label{eq:bny}
f(t,x)-f(t,y) = B(t,x,y)(x-y) \text{ for all }   t\ge 0 \text{ and all }  x,y\in\CC.
\ee
Moreover, for any index~$j$ the following property holds. 
If~$x\in\CC$ with $x_j=0$ then~$f_j(t,x)\ge 0$ for all~$t\ge 0$.
\end{Proposition}
{\sl Proof.} Eq.~\eqref{eq:bny} follows from  the fundamental theorem of calculus with 
\[ 
B(t,x,y) = \int_0^1 J(t,y+s(x-y)) \diff s   \text{ for all } t \ge 0 \text{ and } x,y \in \interior(\CC).
\]
By assumptions (iii) and (iv) on $f$ we conclude that~$B(t,x,y)\in\BB$. 
Moreover, assumption (ii) implies that this formula actually defines~$B$ as a continuous map on   
$[0, \infty) \times \mathbb{R}_{+}^{N}
\times \mathbb{R}_{+}^{N}$ into $\mathcal{B}$. Eq.~\eqref{eq:bny} also   holds   on the closed cone~$\mathbb{R}_{+}^{N}$ due to assumption~(i).
The second claim follows from $f(t,x)=B(t,x,0)x$ and the fact that $B(t,x,0)$ is Metzler.~\IEEEQED

\begin{Lemma}\label{lemma:T1} 
Assume that~$f\in\FF$. Then for every $x_0\in\CC$ the initial value problem $\dot x = f(t,x)$, $x(0)=x_0$,
admits a 
 unique solution $x(\,\cdot\,;x_0):[0,\infty)\to\CC$. Moreover, 
\[ 
H(x) = \sum_{j=1}^N x_j 
\]
is a first integral of the dynamics, i.e., $\frac{d}{dt} H(x(t;x_0))\equiv 0$.
\end{Lemma} 
{\sl Proof.} 
By assumption (i) on $f$ and by Proposition \ref{prop:T1}, the vector field $f$ satisfies the hypotheses of the Picard-Lindel\"of Theorem, however with a domain of definition~$\R^n_+$
 which is not open. Introduce the auxiliary extension 
\[
 \tilde f(t,x) := f(t,|x_1|,\ldots,|x_N|). 
\]
Note that~$ \tilde f:[0,\infty)\times\R^N \to \R^N  $ is well-defined.
The  Picard-Lindel\"of Theorem yields existence and uniqueness   of a maximal solution
$\tilde x(\,\cdot\,;x_0):J_{x_0}\to\R^N$ of the initial value problem $\dot x=\tilde f(t,x)$, $x(0)=x_0$, with 
$J_{x_0} =[0, b_{x_0})$ for some $b_{x_0} >0$ that might be infinite.
Note that~$H$ is a first integral by property~(iv) of $f$, that carries over to $\tilde f$.

On~$\CC$, the solutions $x(\,\cdot\,;x_0)$ and $\tilde x(\,\cdot\,;x_0)$ coincide. 
We now show  that for~$x_0\in\CC$ the solution~$\tilde x(t;x_0) \in\CC$ for all~$t\in J_{x_0}$. 
If $x_0=0$ then $\tilde x(t;x_0)\equiv 0$ by assumption (v) on $f$, so $\tilde x(t;x_0)\in\CC$. 
If $x_0\in\CC\setminus\{0\}$, we argue by contradiction. Assume that there exists~$\tau>0$ such that $\tilde x(\tau;x_0)\in(\R^N\setminus \CC)$.
 For $\eps\in\R$, let
\[ 
\tilde f_\eps:[0,\infty) \times\R^N\to\R^N, \;\; \tilde f_\eps(t,x) := \tilde f(t,x) 
+ \eps \left(\begin{array}{ccc}1 & \cdots & 1\\ \vdots & \vdots & \vdots\\ 1 & \cdots & 1\end{array}\right)x - \eps N x ,
\]
and denote by $\tilde x(\cdot; x_0,\eps)$ the unique maximal solution of $\dot x = \tilde f_\eps(t,x)$, $x(0)=x_0$. Note that by the
definition of $\tilde f_\eps$ the function~$H$ is also a first integral for this dynamical system.
 
Now by the continuous dependence of solutions on parameters 
 there exists $\eps_0>0$ such that $\tilde x(\tau;x_0,\eps_0)\in(\R^N\setminus \CC)$. This implies that there exists a time~$s\in [ 0,\tau)$ such that~$x(s; x_0,\eps_0)$ leaves $\CC$, i.e.
 there exists $k\in\{1,\ldots,N\}$ with $\tilde x_{k}(s;x_0,\eps_0)=0$ and $\dot{\tilde{x}}_{k}(s;x_0,\eps_0)\le 0$. However, this is a
 contradiction, since
\begin{align*}
\dot{\tilde{x}}_{k}(s;x_0,\eps_0)& = (\tilde f_\eps(s,\tilde x(s;x_0,\eps_0)))_{k}\\
& \ge \eps_0 \sum_{j\ne k} \tilde x_j(s; x_0,\eps_0) \\
&= \eps_0 H(x_0) \\&> 0,
\end{align*}
where the first inequality follows from Proposition~\ref{prop:T1} and the definition of~$\tilde f_\eps$,
and the last
equality follows from the fact that~$H$ is also a first integral for $\dot x = \tilde f_\eps(t,x)$.

We have established that $\tilde{x}$ remains in the compact set $\{ z\in\mathbb{R}_{+}^{N}\,|\,H(z) = H(x_0)\}$ and it follows that the solution~$\tilde{x}(t;x_0)$ exists for all $t\ge 0$.
Since the solutions $x(\,\cdot\,;x_0)$ and $\tilde{x}(\,\cdot\,;x_0)$ coincide on $\mathbb{R}_{+}^{N}$,
this completes the proof.~\IEEEQED


\subsection{Linear time-varying systems}\label{ssec:appendix.2}
The properties that are essential in the proofs of our main results are cooperativeness, non-expansiveness, and contractivity of the flow.
As it turns out, it is convenient to first prove these properties for linear time-varying systems. Let
\[ \Ac := \{ A:[0,\infty) \to \BB \,|\, A \mbox{ is continuous}\}.\]
\begin{Lemma}\label{lemma:T2}
Assume that~$A\in\Ac$. Then the initial value problem $\dot x = A(t)x$, $x(0)=x_0 \in\CC$,
 has a unique solution~$x(\,\cdot\,;x_0):[0,\infty) \to \R^N$ that satisfies the following properties: 
\begin{enumerate}
\item[a)] $x(t;x_0) \in\CC$ and $H(x(t;x_0)) = H(x_0)$ for all $t\ge 0$.
\item[b)] If $x_j(t^*;x_0)>0$ for some $j\in\{ 1,\ldots,N\}$ and $t^*\ge 0$
 then $x_j(t;x_0)>0$ for all $t\ge t^*$. 
\item[c)] If $x_0\not =0$ and
 $A(t^*)$ is irreducible for some $t^*\ge 0$  then~$x(t;x_0)\in\interior(\CC)$   for all $t > t^*$.
\end{enumerate}
\end{Lemma}
{\sl Proof.} Existence and uniqueness of the solution are immediate from the linearity and continuity of $A$. 

The proof of~a) follows from Lemma \ref{lemma:T1} and the fact that~$f(t,x) := A(t)x$ belongs to $\FF$ for each
 $A\in\Ac$.

To prove~b), assume that~$x_j(t^*;x_0)>0$. Let~$y(t):= x_j(t;x_0)$. Then $y$ solves the scalar
initial value problem  
\[ 
\dot y = g(t,y), \quad y(t^*)=  x_j(t^*;x_0)>0 \]
where
\[ 
g(t,y) := a_{jj}(t)y + b(t), \quad b(t): = \sum_{k\ne j} a_{jk}(t)x_k(t;x_0) \ge 0.
\]
Thus, letting~$q(t) := \int_{t^*}^t a_{jj}(u) \diff u$ yields 
\begin{align*}
 y(t) &= e^{q(t)} \left(y(t^*) + \int_{t^*}^t e^{-q(s)} b(s) \diff s \right)\\
&\ge e^{q(t)} y(t^*)\\& >0
\end{align*}
for all~$t\geq t^*$. 

To prove property~c) 
first note that irreducibility of~$A(t^*)$ implies that there exists~$\delta>0$ such that
\be\label{eq:itrrpl}
A(t) \text{ is irreducible for all } t\in[t^*,t^*+\delta).
\ee
This follows from the fact that irreducibility  is equivalent to the associated adjacency graph being strongly connected, 
i.e.~certain edges have   positive weights,  
and the continuity of~$A(t)$.

Pick~$x_0\in \R^n_+ \setminus\{0\}$. We consider two cases. 

\noindent {\em Case 1:} $x(t^*;x_0) \in \interior( \CC)$. Then the claim follows from property b).

\noindent {\em Case 2:} $x(t^*;x_0) \in \partial \CC$. Fix $t>t^*$. As $H(x(t^*;x_0)) = H(x_0)>0$, there exists $k\in\{1,\ldots,N-1\}$ such that exactly $k$ entries  of $x(t^*;x_0)$ are positive and the other entries are zero.  Note that by property~b), these~$k$ entries remain positive for all~$t\geq t^*$.
Assume w.l.o.g.\ that the first~$k$ entries  of $x(t^*;x_0)$ are positive.
 Then  in block form
\[ \dot x(t^*;x_0) = \left( \begin{array}{c|c} U & V \\ \hline Y & Z \end{array}\right)\left(\begin{array}{c} x_1(t^*;x_0)\\ \vdots\\ x_k(t^*;x_0)\\ \hline 0\end{array}\right),
\]
with~$U\in\R^{k\times k}$ and~$Z\in\R^{(N-k)\times(N-k)}$.
Since~$A(t^*) $ is Metzler and  irreducible, every entry of~$Y$ is non-negative and at least one 
entry is positive, so      there exists~$j>k$ such that $\dot x_{j}(t^*;x_0)>0$.
Therefore, at least~$k+1$ entries of~$x(t;x_0)$ are positive for~$t > t^*$. 
Now an inductive argument and using~\eqref{eq:itrrpl}  completes the proof.~\IEEEQED

Let
\[ 
\Sc := \left\{ Q\in\R^{N\times N}\,\left|\, q_{jk}\ge 0 \mbox{ for all } j,k=1,\ldots,N \text{ and } \sum_{j=1}^N q_{jk} = 1 \text{ for all }k=1,\ldots,N\right.\right\}
 \]
denote the set of of~$N \times N $ stochastic matrices, and let
\[ 
\Sc^+  :=  \{ Q\in\Sc \, | \, q_{jk}>0 \mbox{ for all } j,k=1,\ldots,N\}
 \]
denote the subset  of stochastic matrices with positive entries.
For $A\in\Ac$ let $\Phi_A:[0,\infty) \to \R^{N\times N}$ be the fundamental matrix of~$\dot x=Ax$, that is,
the solution of 
\[ 
\dot\Phi(t) = A(t)\Phi(t), \quad \Phi(0)=I_N.
\]
Since the columns of $\Phi_A(t)$ are $x(t;e^j)$, where $e^j\in\CC$ denotes the $j$-th canonical unit vector, the next result
 follows from Properties a) and c) in Lemma~\ref{lemma:T2}.

\begin{Corollary}\label{cor:T1} 
Assume that~$A\in\Ac$. Then
\begin{enumerate}
\item[a)] $\Phi_A(t)\in\Sc$ for all $t\ge 0$.
\item[b)] If $A(t^*)$ is irreducible for some $t^* \geq 0$  then $\Phi_A(t)\in\Sc^+$ for all $t>t^*$.
\end{enumerate}
\end{Corollary}

We now use this to prove non-expansiveness with respect to the~$\ell_1$-norm and contractivity in the case of irreducibility.
The first step is to note that 
    stochastic matrices have useful  properties with respect to this norm.
	
\begin{Proposition}\label{prop:T2}
  If $Q\in\Sc$ and $x\in\R^N$ then $|Qx|_1\le |x|_1$.
 If $Q\in\Sc^+$ and $x\in(\R^n\setminus\{0\})$ with $H(x)=0$ then $|Qx|_1<|x|_1$.
\end{Proposition}
{\sl Proof.}  
The first statement  follows from  
\begin{align}\label{eq:sfrt}
 |Qx|_1 &= \sum_{j=1}^N\left|\sum_{k=1}^N Q_{jk}x_k\right| \nonumber\\
& \le \sum_{j=1}^N \sum_{k=1}^N Q_{jk}|x_k|\nonumber\\
& = \sum_{k=1}^N |x_k|  \nonumber\\
&= |x|_1.
\end{align}

To prove the second statement, pick~$x\ne 0$ such that~$H(x)=0$.
Then  there exist $k_1,k_2\in\{1,\ldots,N\}$ such that $x_{k_1}<0<x_{k_2}$. Thus, if~$Q\in \Sc^+$
then for any  $j\in\{1,\ldots,N\}$,
\[ 
\left|\sum_{k=1}^N Q_{jk}x_k\right| < \sum_{k=1}^N Q_{jk}|x_k|
 \]
  because the sum on the left contains both   positive and   negative terms.
	Now arguing
	as in~\eqref{eq:sfrt} completes the proof.~\IEEEQED

\subsection{Non-expansiveness and contractivity}\label{ssec:appendix.3}
Using the   results for time-varying linear systems
we now turn to proving non-expansiveness and contractivity for the nonlinear dynamical  system.

\begin{Theorem}\label{thm:T3} Suppose that $f\in\FF$. Then
\begin{enumerate}
\item[a)] For any $x_0,y_0\in\CC$  the function 
\be\label{eq:nino}
t\mapsto |x(t;x_0)-x(t;y_0)|_1 \text{ is non-increasing on } [0,\infty).
\ee
\item[b)] If there exists  $t^*\ge 0$ such that $J(t^*,x)$ is
 irreducible for all $x\in\Omega$ then  for any $\hat t > t^*$ there exists $M_{\hat t}<1$ such that
\be\label{eq:hatxx}
 |x(\hat t;x_0) - x(\hat t;y_0)|_1 \le M_{\hat t}|x_0-y_0|_1 \quad \text{for all } x_0,y_0\in\Omega.
\ee
\end{enumerate}
\end{Theorem}
{\sl Proof.} Fix $t_1\ge 0$ and~$x_0,y_0\in \Omega$. Let $z(t):= x(t+t_1;x_0)-x(t+t_1;y_0)$. 
By Proposition \ref{prop:T1},   $\dot z(t)= C(t)z(t)$, with~$z(0)=z_0:= x(t_1;x_0)-x(t_1;y_0)$, and $C\in\Ac$ given by 
\[
 C(t) := B(t+t_1,x(t+t_1;x_0),x(t+t_1;y_0))
 \]
for all $t\ge 0$. Hence, $z(t) = \Phi_C(t)z_0$ for all $t\ge 0$. 

To prove~\eqref{eq:nino}, pick $t_2\ge t_1$. By Corollary \ref{cor:T1}a), $\Phi_C(t_2-t_1)\in\Sc$  and Proposition~\ref{prop:T2}  yields 
\begin{align*}
 |x(t_2;x_0)-x(t_2;y_0)|_1 &= |z(t_2-t_1)|_1 \\&= |\Phi_C(t_2-t_1)z_0|_1 \\& \le |z_0|_1 \\&= |x(t_1;x_0) - x(t_1;y_0)|_1, 
\end{align*}
and this   proves~\eqref{eq:nino}. 

	To prove~\eqref{eq:hatxx},
let
		\[
			\PSET:=\{x\in\R^N  \,|\,   H(x)=0 \text{ and }  |x|_1=1\}. 
		\]
		Note that if~$y,z\in \Omega$ with~$y\not = z$ then $  \frac{y-z}{|y-z|_1}    \in \PSET$.
		
		Pick~$\hat t>t^*$ and~$x_0,y_0\in\Omega$.
Eq.~\eqref{eq:hatxx} clearly holds if~$x_0=y_0$, so we may assume that~$x_0\not =y_0$.
Then
\begin{align*} |x(\hat t;x_0)-x(\hat t;y_0)|_1    
 &=   |\Phi_{C }(\hat t)(x_0-y_0)|_1\\
& =     m\left(x_0,y_0,\frac{x_0-y_0}{|x_0-y_0|_1}\right )|x_0-y_0|_1 ,
\end{align*}
where
$
 m:\Omega\times\Omega\times\PSET \to \R
$
is defined by
$
 m(x_0,y_0,v):=|\Phi_{C }(\hat t)v|_1
$.
Let
\[
					M_{\hat t}:=\max \{ m(x,y,v) \; | \; x\in \Omega, y \in \Omega, v \in \PSET   \}.  
\]
This is well-defined, as 
the matrix~$C(t) = C_{x_0,y_0}(t) = B(t,x(t;x_0),x(t;y_0))$ depends 
 continuously on $x_0,y_0\in\Omega$ for all $t\ge 0$ (see   Proposition~\ref{prop:T1})  
and thus~$\Phi_{C_{x_0,y_0}}(\hat t)$ is also continuous in~$x_0,y_0$. 
We conclude that
\begin{align*} |x(\hat t;x_0)-x(\hat t;y_0)|_1    
 & \le M_{\hat t}|x_0-y_0|_1.
\end{align*}
Thus, to complete the proof 
 we only need to show that $M_{\hat t}<1$. 
To prove this, 
denote $x^* := x(t^*;x_0)$, $y^* := x(t^*;y_0)$. Then
\[
 C_{x_0,y_0}(t^*) = B(t^*,x^*,y^*) = \int_0^1 J(t^*,y^* + s(x^*-y^*)) \diff s.
\]
Since~$J$ is Metzler (i.e. all its off diagonal elements are non-negative) and, by assumption, $J(t^*,z)$ is irreducible for all~$z\in\Omega$,
we conclude that~$C_{x_0,y_0}(t^*)$ is also irreducible.    Corollary~\ref{cor:T1} implies that 
$\Phi_{C_{x_0,y_0}}(\hat t)\in\Sc^+$.
Picking a maximizer $(x_0, y_0, v_0) \in \Omega \times \Omega \times \mathcal{P}$ of $m$, i.e.~$M_{\hat{t}} = m(x_0, y_0, v_0)$, it follows from Proposition \ref{prop:T2}  that~$M_{\hat{t}} < |v_0|_1 = 1$.~\IEEEQED

\subsection{Proof of Theorem~\ref{thm:T2}}\label{ssec:appendix.4}
We can now   prove Theorem~\ref{thm:T2}. We note that the proof proceeds without the explicit use of the cooperative behavior of dynamical systems 
(though we will use this property for the proof of Theorem~\ref{thm:T1}; see Subsections~\ref{sec:coop},~\ref{ssec:appendix.6}   below). 

Note that for $f\in\FF_T$ a solution $\dot\gamma = f(t,\gamma)$, $\gamma:[0,\infty)\to\Omega$, is $T$-periodic if and only if~$x(T;\gamma(0))=\gamma(0)$. Thus, consider  the period map~$P_T:\CC\to\CC$ defined by 
\begin{equation}  
 P_T(a) := x(T;a). 
\label{eq:PT}
\end{equation}
In other words,~$P_T(a) $   is the value of~$x(T)$ for the initial condition~$x(0)=a$. 
Observe that $P_T(\Omega)\subseteq \Omega$ (as $H$ is a first integral of $\dot x = f(t,x)$). Moreover, for $f\in(\FF_T\cap \FF^\Omega_{\text{irr}})$ there exists $t^*\in[0,T)$ such that
$J(t^*,x)$ is irreducible for all~$x\in\Omega$.   Then~$T>t^*$, so
 Theorem~\ref{thm:T3}b) implies that $P_T$ is Lipschitz on the closed set $\Omega$ with Lipschitz constant~$M_T<1$.
The Banach fixed point theorem implies that~$P_T$ has a  unique fixed point in~$\Omega$, that is,
 there exists a   unique $T$-periodic function~$\gamma:\R\to\Omega$ that solves~$\dot\gamma = f(t,\gamma)$. 
Fix~$\alpha>0$ such that
\be\label{eq:bounjk}
\max\{1/2,M_T\} \le e^{-\alpha T} .
\ee
Pick~$x_0\in\Omega$ and $t\ge 0$, and let
  $k\in\N_0$ be such that $kT\le t \le (k+1)T$. Then   Theorem~\ref{thm:T3}  yields
\begin{align*}
 |x(t;x_0)-\gamma(t)|_1 &\le |x(kT;x_0)-\gamma(kT)|_1 \\
&\le (M_T)^k|x_0-\gamma(0)|_1\\
& \le e^{-\alpha Tk}|x_0-\gamma(0)|_1 \\
& = e^{-\alpha T (k+1)} e^{ \alpha T  } |x_0-\gamma(0)|_1 \\
&\le e^{\alpha T} e^{-\alpha t} |x_0-\gamma(0)|_1\\
&\le 2 e^{-\alpha t} |x_0-\gamma(0)|_1.
 \end{align*}
where the last inequality follows from~\eqref{eq:bounjk}.
  Thus, every solution~$x(t;x_0)$
 converges to the unique periodic solution~$\gamma(t)$ at an exponential rate.~\IEEEQED

\subsection{Cooperative behavior}\label{sec:coop}
For the proof of Theorem \ref{thm:T1} we use some elegant topological ideas from \cite{mono_periodic_96,Dancer1991}, which are based on the cooperative behavior of the dynamical system generated by a vector field~$f\in\FF$.  In order to formulate this concept we introduce some more notation. 

Let $p,q\in\R^N$, and let~$A$ be a bounded  non-empty subset of~$\R^N$. Then we write
\begin{enumerate}
\item[a)] $p\le q$ $:\Leftrightarrow$ $p_j\le q_j$ for all $j\in\{1,\ldots,N\}$,
\item[b)] $[p,q] := \{x\in\R^N\,|\, p\le x\le q\}$,
\item[c)] $\inf A := c\in\R^N$ with $c_j:=\inf\{ a_j\,|\, a\in A\}$,\\
$\sup A := d\in\R^N$ with $d_j:=\sup\{ a_j\,|\, a\in A\}$,
\item[d)] $p\le A$ $ \Leftrightarrow$ $p\le a$ for all $a\in A$\\
$q\ge A$ $ \Leftrightarrow$ $q\ge a$ for all $a\in A$,
\end{enumerate}
It is straightforward to verify that $\inf A\le A\le \sup A$ and that for all $x,y\in\R^N$ with $x\le A$, $y\ge A$ we have $x\le \inf A$ and $y\ge \sup A$.

The following theorem summarizes the monotone  behavior with respect to the order~$\leq $.

\begin{Theorem}\label{thm:T4}
Let $f\in\FF$ and $x_0,y_0\in \CC$ with $x_0\le y_0$. Then $x(t;x_0)\le x(t;y_0)$ for all $t\ge 0$. If, in addition, $(x_0)_j<(y_0)_j$ for some $j\in\{1,\ldots,N\}$, then $x_j(t;x_0)< x_j(t;y_0)$ for all $t \ge 0$.
\end{Theorem}
{\sl Proof.} We use again that $z(t):=x(t;y_0)-x(t;x_0)$ solves $\dot z = C(t)z$, $z(0)=z_0:=y_0-x_0\in\CC$ with $C(t): = B(t,x(t;y_0),x(t;x_0))\in\BB$ and $C\in\Ac$. The claim then follows from statements a) and b) of Lemma \ref{lemma:T2}.~\IEEEQED

\subsection{$\omega$-Limit sets of $P_T$ and proof of Theorem \ref{thm:T1}}\label{ssec:appendix.6}
The concept of $\omega$-limit sets for the discrete time dynamical system induced by the period map $P_T:\Omega\to\Omega$ from \eqref{eq:PT} is pivotal for the proof of Theorem \ref{thm:T1}. For $a\in\Omega$   this set is     
\[ \omega_T(a) := \{x\in\Omega\,|\, \mbox{there is a sequence } n_k\to\infty \mbox{ with } \lim_{k\to\infty} P_T^{n_k}(a) = x\}.\]
We first state a few standard facts about this set.

\begin{Proposition}\label{prop:T3} Let $f\in\FF$.  Pick~$T>0$. Then for all $a\in\Omega$ we have
\begin{enumerate}
\item[a)] $\omega_T(a)$ is a closed, nonempty subset of $\Omega$;
\item[b)] $P_T(\omega_T(a)) = \omega_T(a)$;
\item[c)] $\omega_T(b)\subseteq \omega_T(a)$ for all $b\in\omega_T(a)$.
\end{enumerate}
\end{Proposition}
{\sl Proof.} Statements a) and b) follow from \cite[Eq.\ (4.7.2) and Lemma 4.7.4]{MiWH01} because $P^k_T(a)$ evolves in the compact set $\Omega$.


To prove  c), pick~$b\in\omega_T(a)$. Property  b) implies that~$P^{n_k}_T(b)\in\omega_T(a)$ for all $n_k\in\N$, and since~$\omega_T(a)$  is closed
   $\lim_{k\to\infty} P_T^{n_k}(b)\in\omega_T(a)$.~\IEEEQED

The following lemma, that is proved in the subsequent subsection, provides all that is needed to prove Theorem~\ref{thm:T1}.

\begin{Lemma}\label{lemma:T4A}
Let $f \in \mathcal{F}_T$.  Then for every $x \in \Omega$ its limit set $\omega_T(x)$ contains a fixed point of $P_T$.
\end{Lemma}
{\em Proof of Theorem \ref{thm:T1}.}\;
 Pick $x_0 \in \Omega$ and denote by $z \in \omega_T(x_0)$ the fixed point of $P_T$ that exists 
according to Lemma~\ref{lemma:T4A}.
Since $z$ is a fixed point and since $f \in \mathcal{F}_T$ is $T$-periodic, the solution $\gamma(t) := x(t; z)$ is also $T$-periodic. As $z \in \omega_T(x_0)$
there exists a subsequence $P_T^{n_k}(x_0) \to z$ as $k \to \infty$. For any $j \in \mathbb{N}$ and $t \ge n_j T$ we have by Theorem \ref{thm:T3}a), 
\[ |x(t;x_0)-\gamma(t)|_1 \le |x(Tn_j;x_0)-\gamma(Tn_j)|_1 = |P_T^{n_j}(x_0)-z|_1  .
\]
As~$t\to \infty$, we can take~$j\to \infty$ and this yields
\[
 \lim_{t\to\infty} |x(t;x_0)-\gamma(t)|_1 =0,
\]
proving Theorem \ref{thm:T1}.~\IEEEQED

Note that the proof  of Theorem~\ref{thm:T1} shows that for any~$x_0\in \Omega$
the~$\omega$-limit set $\omega_T(x_0)$ cannot contain more than one fixed point of~$P_T$.
Thus, the statement in Lemma~\ref{lemma:T4A} can actually be strengthened  
to~$\omega_T(x_0)$ contains exactly one fixed point of $P_T$.

The next subsection contains the proof of the crucial Lemma~\ref{lemma:T4A}.
We have adapted the ideas presented by J. Ji-Fa in 
\cite{mono_periodic_96} to our setting which led to somewhat simplified arguments. In particular, we can replace
\cite[Proposition 1]{Dancer1991} that is used in the proof of Lemma 3.2 of \cite{mono_periodic_96} 
by a standard application of Brouwer's fixed point theorem.
Indeed, the claim of Lemma~\ref{lemma:T4A} is the existence of a fixed point of the map $P_T$ in~$\omega_T(x)$.  
 We show   that $\omega_T(x)$ contains an element $y$ so that $\omega_T(y)$ is a singleton, say,~$\omega_T(y)=\{ z \}$. Then $z \in \omega_T(x)$ by Proposition \ref{prop:T3}c) and $z$ is a fixed point by Proposition \ref{prop:T3}b).

\subsection{Proof of Lemma~\ref{lemma:T4A}}\label{ssec:appendix.7}

We use the following notation.
For $y\in\Omega$, let
\begin{align*}
p(y)&:= \inf\omega_T(y),\\
q(y) &:= \sup\omega_T(y).
\end{align*}  
Note that since $\omega_T(y)\subseteq \Omega\subset\CC$, $p(y)$ and $q(y)$ exist in $\CC$. 
Moreover, $p(y) \leq q(y)$ and   $\omega_T(y)$ is a singleton if and only if $p(y) = q(y)$. 
The existence of the desired element $y$ in $\omega_T(x)$ is established
by contradiction. Suppose that no such $y$ exists. Denote by $y_0$ an element in $\omega_T(x)$ that \emph{minimizes} the number of coordinates $j$
for which $(p(y))_j$ and $(q(y))_j$ differ. We then show that there exists~$z_0 \in \omega_T(x)$ for which $p(z_0)$ and $q(z_0)$ 
differ in a smaller number of coordinates than $p(y_0)$ and $q(y_0)$. Part c) of Lemma \ref{lemma:T3} below states a fact that is essential for the
construction of $z_0$. What is also crucial for the proof is the observation that $p(y)$ and $q(y)$ are fixed points of $P_T$ which is formulated in 
Lemma~\ref{lemma:T3}a). Its short proof demonstrates why \emph{ monotone dynamical systems 
admitting a first integral with a positive gradient}
are  special.

For $y \in \Omega$, let
\begin{align} \label{eq:defgyp}
J_y&:=\{ j\in\{1,\ldots,N\}\,|\, (p(y))_j\ne (q(y))_j\},
\end{align}
i.e. the set of indices for which $(p(y))_j$ and $(q(y))_j$ differ.

\begin{Lemma}\label{lemma:T3}
Let $f\in\FF_T$. Then for any~$y\in\Omega$,  
\begin{enumerate}
\item[a)] $p(y)$, $q(y)$ are fixed points of $P_T$.
\item[b)] $P_T([p(y),q(y)]) \subseteq [p(y),q(y)]$.
\item[c)] If $p(y)\ne q(y)$
then for any $z\in\omega_T(y)$ there exists a~$j\in J_y$ such that $z_j=(p(y))_j$.
\end{enumerate}
\end{Lemma}

To explain  
property c), we introduce more notation. 
For $v,w\in\R^N$ let 
\[
\Delta(v,w):=\#\{j\in\{1,\ldots,N\}\,|\, v_j \ne w_j\}.
\]
Consider the case~$p(y)\ne q(y)$. For any~$j \in \{1,\dots,N\}\setminus J_y$ we have~$p_j(y)=q_j(y)$, so~$z_j=p_j=q_j$.
For any~$j \in   J_y$ we have~$p_j(y)<q_j(y)$, and property c) implies that there exists at least one such index such that~$p_j(y)=z_j$. 
We conclude that
\be\label{eq:meanc}
										\Delta(z,p(y))<\Delta(q(y),p(y)).
\ee

{\sl Proof of Lemma~\ref{lemma:T3}.}
 To simplify the notation we write from hereon~$p,q$ for~$p(y),q(y)$. Pick~$z\in\omega_T(y)$.
  As $p \le z$, Theorem~\ref{thm:T4} implies that~$P_T(p )\le P_T(z)$. Thus $P_T(p )\le \omega_T(y)$ by Proposition \ref{prop:T3}b), and consequently $P_T(p )\le p $. Since $H(P_T(p ))=H(p )$, it follows that $P_T(p )=p $. The proof that $P_T(q )=q $ proceeds analogously. This proves claim~a). 

Claim~b) follows directly from~a) and Theorem~\ref{thm:T4}.

We prove c) by contradiction. Assume that  there exists $z\in\omega_T(y)$ such that $z_j> p_j$ for all $j\in J_y$. Set 
\be\label{Eq:defeps}
\eps:=\min\{z_j-p_j\,|\, j\in J_y\}.
\ee
 Then $\eps>0$. Define
\be\label{eq:defggg}
\Gamma:=\{x\in[p,q]\,|\, H(x) = H(p) + \frac\eps2\}.
\ee
Note that for any~$j \in \{1,\dots,N\} \setminus J_y$ we have  $ p_j=q_j$ so if~$x\in \Gamma$ then~$x_j=p_j=q_j$.  
 The set $\Gamma $ is not empty, as
 $H(q)\ge H(z)\ge H(p)+\eps$.
 $\Gamma$ is also
compact and convex. Statement b) and the fact that $H$ is a first integral imply  that~$P_T(\Gamma)\subseteq \Gamma$.
The Brouwer fixed point theorem thus yields the existence
 of a fixed point of~$P_T$ in~$\Gamma$, that is, there exists~$x^*\in\Gamma$ such that~$P_T(x^*)=x^*$.

Since $z\in\omega_T(y)$ there exists $n^*\in\N$ such that~$|P_T^{n^*}(y)-z|_1 < \frac{\eps}{3N}$. Define $y^*$ by 
\[ 
y^*_j := \left\{ \begin{array}{ll} (P_T^{n^*}(y))_j, &   j\in J_y\\ z_j=p_j=q_j, & \text{otherwise.}\end{array}\right. 
\]
Since $|P_T^{n^*}(y)-y^*|_1+|y^*-z|_1 \leq |P_T^{n^*}(y)-z|_1$, we have $|P_T^{n^*}(y)-y^*|_1<\frac{\eps}{3N}$ and 
\be\label{eq:poli}
|y^*-z|_1<\frac{\eps}{3N}. 
\ee
Observe that 
\[ 
x_j^* = p_j = z_j = y_j^* \quad \text{ for } j\in\{1,\ldots,N\}\setminus J_y ,
\]
and
\[ 
x_j^* \le p_j + \frac\eps2\le z_j - \frac\eps2 <  y_j^* \quad \text{ for } j\in J_y, 
\]
where the first inequality follows from~\eqref{eq:defggg},
 the second    from~\eqref{Eq:defeps},
and the third from~\eqref{eq:poli}.

Summarizing, $x^*\le y^*$. Since $P_T(x^*)=x^*$, Theorem~\ref{thm:T4} implies that 
$x^*\le P_T^k(y^*)$ for all $k\ge 0$. 
Since $\sum_j(x_j^*-p_j) = \frac\eps2$, there exists
  $j_0\in\{1,\ldots,N\}$ such that  
$x_{j_0}^* \ge p_{j_0} + \frac{\eps}{2N}$. 
Then $(P_T^k(y^*))_{j_0}\ge p_{j_0}+\frac{\eps}{2N}$ for all $k\ge 0$.  

Since $P_T$ is non-expansive by Theorem \ref{thm:T3}a), 
we have in addition $|P_T^k(y^*)-P_T^{k+n^*}(y)|_1 \le |y^*-P_T^{n^*}(y)|_1 < \frac{\eps}{3N}$ for all $k\ge 0$. 
Hence for all~$m\ge n^*$,
\[
(P_T^m(y))_{j_0}\ge p_{j_0} + \frac\eps{2N}-\frac\eps{3N} = p_{j_0} + \frac{\eps}{6N}  .
\]
 We conclude that 
   any $r\in\omega_T(y)$ satisfies~$r_{j_0}\ge p_{j_0} + \frac\eps{6N}$, and thus
\[ 
p_{j_0} = \inf\{ r_{j_0}\,|\, r\in\omega_T(y)\} \ge p_{j_0} + \frac\eps{6N}, 
\]
and this contradiction proves~c).~\IEEEQED

We can now prove the crucial  Lemma \ref{lemma:T4A}.

{\sl Proof of Lemma \ref{lemma:T4A}.}\; 
As noted above  we need to show that there exists~$y \in \omega_T(x)$ such that
$p(y)=q(y)$. To do this,   for $y\in\Omega$ let
\[
 m(y):=\Delta(p(y),q(y)),
\]
and for~$x\in \Omega$, let 
\[
\alpha(x ):= \min\{ m(y)\,|\, y\in\omega_T(x)\} .
\]
It suffices to show that $\alpha(x) = 0$ for all~$x\in\Omega$. We achieve this by contradiction.
%
%
Assume that there exists~$x\in\Omega$ for which~$\alpha(x)>0$. Then there exists~$y_0\in\omega_T(x)$ with $\alpha(x) = m(y_0) >0$. 
Let~$\tilde M := \max\{\Delta(z,p(y_0))\,|\, z\in\omega_T(y_0)\}$. Then~\eqref{eq:meanc} yields 
\be\label{eq:con1}
\tilde M <\alpha(x).
\ee
 Choose $z_0\in\omega_T(y_0)$ such that $\Delta(z_0,p(y_0))=\tilde M$. 
We now  show that~$m(z_0)\le \tilde M$. To this end, define 
\[
J:=\{j\in\{1,\ldots.N\}\,|\, (z_0)_j> p_j(y_0)\}.
\]
Note that $\tilde M = \# J$. 
Since $z_0\in[p(y_0),q(y_0)]$, we have $J\subseteq J_{y_0}$ and by Lemma~\ref{lemma:T3} we have $J\not =  J_{y_0}$.
 Consider the sequence   $z(k) := P_T^k(z_0) \in \omega_T(y_0)$. By the second statement of Theorem \ref{thm:T4} we 
have $z_j(k) > p_j(y_0)$ for all $j\in J$ and all $k\in\N$ (recall that $p(y_0)$ is a fixed point). 
Hence $\Delta(z(k),p(y_0))\ge \# J = \tilde M$.
 Since $z(k) \in\omega_T(y_0)$, we have by the definition of $\tilde M$ also $\Delta(z(k),p(y_0))\le \tilde M$ and,
 therefore, $\Delta(z(k),p(y_0)) = \tilde M$, and 
\[ 
J=\{ j\in\{1,\ldots,N\}\,|\, z_j(k) >p_j (y_0)\}  \quad \text{ for all } k\in\N.
\]
Thus $z_j(k) = p_j (y_0)$ for all $j\in\{1,\ldots,N\}\setminus J$, so 
\[ 
\omega_T(z_0) \subseteq \{ v\in\R^N\,|\, v_j = p_j(y_0) \text{ for all } j \in \{1,\ldots,N\}\setminus J\}.
\]
This implies that $p_j(z_0)=q_j(z_0)  $ for all $j\in\{1,\ldots,N\}\setminus J$,
 and thus
$m(z_0) = \Delta(p(z_0),q(z_0)) \le \# J = \tilde M$.
Combining this with the fact that~$z_0\in\omega_T(y_0)\subseteq \omega_T(x)$ and~\eqref{eq:con1} yields
\[
 \alpha(x) \le m(z_0)\le \tilde M < \alpha(x),
\]
and this contradiction completes the proof of Lemma~\ref{lemma:T4A}.~\IEEEQED

%
%

\newpage




\begin{thebibliography}{10}
\providecommand{\url}[1]{#1}
\csname url@rmstyle\endcsname
\providecommand{\newblock}{\relax}
\providecommand{\bibinfo}[2]{#2}
\providecommand\BIBentrySTDinterwordspacing{\spaceskip=0pt\relax}
\providecommand\BIBentryALTinterwordstretchfactor{4}
\providecommand\BIBentryALTinterwordspacing{\spaceskip=\fontdimen2\font plus
\BIBentryALTinterwordstretchfactor\fontdimen3\font minus
  \fontdimen4\font\relax}
\providecommand\BIBforeignlanguage[2]{{%
\expandafter\ifx\csname l@#1\endcsname\relax
\typeout{** WARNING: IEEEtran.bst: No hyphenation pattern has been}%
\typeout{** loaded for the language `#1'. Using the pattern for}%
\typeout{** the default language instead.}%
\else
\language=\csname l@#1\endcsname
\fi
#2}}

\bibitem{GILLESPIE1992404}
D.~T. Gillespie, ``A rigorous derivation of the chemical master equation,''
  \emph{Physica A: Statistical Mechanics and its Applications}, vol. 188,
  no.~1, pp. 404--425, 1992.

\bibitem{meq_book}
G.~Haag, \emph{Modelling with the Master Equation: Solution Methods and
  Applications in Social and Natural Sciences}.\hskip 1em plus 0.5em minus
  0.4em\relax Cham, Switzerland: Springer International Publishing, 2017.

\bibitem{stokamp}
N.~G. {Van Kampen}, \emph{Stochastic Processes in Physics and Chemistry},
  3rd~ed.\hskip 1em plus 0.5em minus 0.4em\relax Amsterdam: Elsevier, 2007.

\bibitem{adven_stoch}
S.~I. Resnick, \emph{Adventures in Stochastic Processes}.\hskip 1em plus 0.5em
  minus 0.4em\relax Boston, MA: Birkhauser, 2002.

\bibitem{book_mastereq}
R.~Toral and P.~Colet, \emph{Stochastic Numerical Methods}.\hskip 1em plus
  0.5em minus 0.4em\relax Weinheim, Germany: Wiley, 2014.

\bibitem{TASEP_book}
A.~Schadschneider, D.~Chowdhury, and K.~Nishinari, \emph{Stochastic Transport
  in Complex Systems: From Molecules to Vehicles}.\hskip 1em plus 0.5em minus
  0.4em\relax Elsevier, 2011.

\bibitem{krug2016}
J.~Krug, ``Nonequilibrium stationary states as products of matrices,'' \emph{J.
  Phys. A: Math. Theor.}, vol.~49, p. 421002, 2016.

\bibitem{nadler_schulten1986}
W.~Nadler and K.~Schulten, ``Generalized moment expansion for observables of
  stochastic processes in dimensions $d>1$: Application to {Mossbauer} spectra
  of proteins,'' \emph{J. Chem. Phys.}, vol.~84, no.~7, pp. 4015--4025, 1986.

\bibitem{entrain2011}
G.~Russo, M.~di~Bernardo, and E.~D. Sontag, ``Global entrainment of
  transcriptional systems to periodic inputs,'' \emph{{PLOS Computational
  Biology}}, vol.~6, p. e1000739, 2010.

\bibitem{Keith1984}
W.~L. Keith and R.~H. Rand, ``1:1 and 2:1 phase entrainment in a system of two
  coupled limit cycle oscillators,'' \emph{J. Math. Bio.}, vol.~20, no.~2, pp.
  133--152, 1984.

\bibitem{epidemics_2006}
N.~C. Grassly and C.~Fraser, ``Seasonal infectious disease epidemiology,''
  \emph{Proc. Royal Society B: Biological Sciences}, vol. 273, p. 2541–2550,
  2006.

\bibitem{sync_traffic}
R.~Donner, ``Emergence of synchronization in transportation networks with
  biologically inspired decentralized control,'' in \emph{Recent Advances in
  Nonlinear Dynamics and Synchronization}, ser. Studies in Computational
  Intelligence, K.~Kyamakya, H.~Unger, J.~C. Chedjou, N.~F. Rulkov, and Z.~Li,
  Eds.\hskip 1em plus 0.5em minus 0.4em\relax Berlin Heidelberg:
  Springer-Verlag, 2009, vol. 254.

\bibitem{dorfler16}
D.~{Gro{\ss}}, C.~{Arghir}, and F.~{D{\"o}rfler}, ``{On the steady-state
  behavior of a nonlinear power system model},'' \emph{ArXiv e-prints}, 2016.

\bibitem{sontag_cotraction_tutorial}
Z.~Aminzare and E.~D. Sontag, ``Contraction methods for nonlinear systems: A
  brief introduction and some open problems,'' in \emph{{Proc.\ 53rd IEEE Conf.
  on Decision and Control}}, Los Angeles, CA, 2014, pp. 3835--3847.

\bibitem{LOHMILLER1998683}
W.~Lohmiller and J.-J.~E. Slotine, ``On contraction analysis for non-linear
  systems,'' \emph{Automatica}, vol.~34, pp. 683--696, 1998.

\bibitem{coogan_margaliot}
\BIBentryALTinterwordspacing
M.~Margaliot and S.~Coogan, ``Approximating the frequency response of
  contractive systems,'' \emph{ArXiv e-prints}, 2017. [Online]. Available:
  \url{http://adsabs.harvard.edu/abs/2017arXiv170206576M}
\BIBentrySTDinterwordspacing

\bibitem{zadeh_book_1963}
L.~A. Zadeh and C.~A. Desoer, \emph{Linear System Theory}.\hskip 1em plus 0.5em
  minus 0.4em\relax McGraw-Hill, 1963.

\bibitem{hlsmith}
H.~L. Smith, \emph{Monotone Dynamical Systems: An Introduction to the Theory of
  Competitive and Cooperative Systems}, ser. Mathematical Surveys and
  Monographs.\hskip 1em plus 0.5em minus 0.4em\relax Providence, RI: Amer.
  Math. Soc., 1995, vol.~41.

\bibitem{Mierc1991}
J.~Mierczynski, ``A class of strongly cooperative systems without
  compactness,'' \emph{Colloq. Math.}, vol.~62, pp. 43--47, 1991.

\bibitem{cast_book}
M.~Margaliot, T.~Tuller, and E.~D. Sontag, ``Checkable conditions for
  contraction after small transients in time and amplitude,'' in \emph{Feedback
  Stabilization of Controlled Dynamical Systems: In Honor of {Laurent}
  {Praly}}, N.~Petit, Ed.\hskip 1em plus 0.5em minus 0.4em\relax Cham,
  Switzerland: Springer International Publishing, 2017, pp. 279--305.

\bibitem{MacDonald1968}
C.~T. MacDonald, J.~H. Gibbs, and A.~C. Pipkin, ``Kinetics of biopolymerization
  on nucleic acid templates,'' \emph{Biopolymers}, vol.~6, pp. 1--25, 1968.

\bibitem{TASEP_tutorial_2011}
R.~Zia, J.~Dong, and B.~Schmittmann, ``Modeling translation in protein
  synthesis with \mbox{TASEP}: A tutorial and recent developments,'' \emph{J.
  Statistical Physics}, vol. 144, pp. 405--428, 2011.

\bibitem{Shaw2003}
L.~B. Shaw, R.~K.~P. Zia, and K.~H. Lee, ``Totally asymmetric exclusion process
  with extended objects: a model for protein synthesis,'' \emph{Phys. Rev. E},
  vol.~68, p. 021910, 2003.

\bibitem{kriecherbauer_krug2010}
T.~Kriecherbauer and J.~Krug, ``A pedestrian's view on interacting particle
  systems, \mbox{KPZ} universality, and random matrices,'' \emph{J. Phys. A:
  Math. Theor.}, vol.~43, p. 403001, 2010.

\bibitem{traffic_TASEP_99}
D.~Chowdhury, L.~Santen, and A.~Schadschneider, ``Vehicular traffic: A system
  of interacting particles driven far from equilibrium,'' \emph{Curr. Sci.},
  vol.~77, pp. 411--419, 1999.

\bibitem{mat_ana_sec_ed}
R.~A. Horn and C.~R. Johnson, \emph{Matrix Analysis}, 2nd~ed.\hskip 1em plus
  0.5em minus 0.4em\relax Cambridge University Press, 2013.

\bibitem{PhysRevE.78.011122}
V.~Popkov, M.~Salerno, and G.~M. Sch\"utz, ``Asymmetric simple exclusion
  process with periodic boundary driving,'' \emph{Phys. Rev. E}, vol.~78, p.
  011122, 2008.

\bibitem{PhysRevE.93.012123}
A.~F. Yesil and M.~C. Yalabik, ``Dynamical phase transitions in totally
  asymmetric simple exclusion processes with two types of particles under
  periodically driven boundary conditions,'' \emph{Phys. Rev. E}, vol.~93, p.
  012123, 2016.

\bibitem{PhysRevE.83.031115}
U.~Basu, D.~Chaudhuri, and P.~K. Mohanty, ``Bimodal response in periodically
  driven diffusive systems,'' \emph{Phys. Rev. E}, vol.~83, p. 031115, 2011.

\bibitem{RFM_entrain}
M.~Margaliot, E.~D. Sontag, and T.~Tuller, ``Entrainment to periodic initiation
  and transition rates in a computational model for gene translation,''
  \emph{PLoS ONE}, vol.~9, no.~5, p. e96039, 2014.

\bibitem{RFM_concave}
G.~Poker, Y.~Zarai, M.~Margaliot, and T.~Tuller, ``Maximizing protein
  translation rate in the nonhomogeneous ribosome flow model: a convex
  optimization approach,'' \emph{J. Royal Society Interface}, vol.~11, no. 100,
  2014.

\bibitem{RFM_model_compete_J}
A.~Raveh, M.~Margaliot, E.~D. Sontag, and T.~Tuller, ``A model for competition
  for ribosomes in the cell,'' \emph{J. Royal Society Interface}, vol.~13, no.
  116, 2016.

\bibitem{RFM_sense}
G.~Poker, M.~Margaliot, and T.~Tuller, ``Sensitivity of {mRNA} translation,''
  \emph{Sci. Rep.}, vol.~5, p. 12795, 2015.

\bibitem{10.1371/journal.pone.0166481}
Y.~Zarai, M.~Margaliot, and T.~Tuller, ``On the ribosomal density that
  maximizes protein translation rate,'' \emph{PLOS ONE}, vol.~11, no.~11, pp.
  1--26, 11 2016.

\bibitem{solvers_guide}
R.~A. Blythe and M.~R. Evans, ``Nonequilibrium steady states of matrix-product
  form: a solver's guide,'' \emph{J. Phys. A: Math. Theor.}, vol.~40, no.~46,
  pp. R333--R441, 2007.

\bibitem{DerE97}
B.~Derrida and M.~R. Evans, ``The asymmetric exclusion model: exact results
  through a matrix approach,'' in \emph{Nonequilibrium Statistical Mechanics in
  One Dimension}, V.~Privman, Ed.\hskip 1em plus 0.5em minus 0.4em\relax
  Cambridge, UK: Cambridge University Press, 1997, pp. 277--304.

\bibitem{nasell}
I.~N\r{a}sel, \emph{Extinction and Quasi-Stationarity in the Stochastic
  Logistic {SIS} Model}, ser. Lecture Notes in Mathematics.\hskip 1em plus
  0.5em minus 0.4em\relax Berlin, Germany: Springer, 2011, vol. 2022.

\bibitem{Baca2015}
N.~Baca{\"e}r, ``On the stochastic {SIS} epidemic model in a periodic
  environment,'' \emph{J. Math. Bio.}, vol.~71, no.~2, pp. 491--511, 2015.

\bibitem{Soper1929}
H.~E. Soper, ``The interpretation of periodicity in disease prevalence,''
  \emph{J. Royal Statistical Society}, vol.~92, no.~1, pp. 34--73, 1929.

\bibitem{krug1991}
J.~Krug, ``Boundary-induced phase transitions in driven diffusive systems,''
  \emph{Phys. Rev. Lett.}, vol.~67, pp. 1882--1885, 1991.

\bibitem{peri_gross_subs}
F.~Nakajima, ``Periodic time dependent gross-substitute systems,'' \emph{SIAM
  J. Appl. Math.}, vol.~36, no.~3, pp. 421--427, 1979.

\bibitem{Dancer1991}
E.~N. Dancer and P.~Hess, ``Stability of fixed points for order-preserving
  discrete-time dynamical systems,'' \emph{J. reine angew. Math.}, vol. 419,
  pp. 125--139, 1991.

\bibitem{mono_periodic_96}
J.~Ji-Fa, ``Periodic monotone systems with an invariant function,'' \emph{SIAM
  J. Math. Anal.}, vol.~27, pp. 1738--1744, 1996.

\bibitem{MiWH01}
A.~N. Michel, K.~Wang, and B.~Hu, \emph{Qualitative Theory of Dynamical
  Systems}, 2nd~ed., ser. Monographs and Textbooks in Pure and Applied
  Mathematics.\hskip 1em plus 0.5em minus 0.4em\relax New York: Marcel Dekker,
  2001, vol. 239.

\end{thebibliography}

\end{document}